\documentclass[a4paper,11pt]{article}
\pdfoutput=1 

\usepackage{jcappub} 

\usepackage[T1]{fontenc} 
\usepackage{amsthm}
\usepackage{mathtools}
\usepackage{orcidlink}
\DeclareMathOperator{\cov}{cov}

\title{\boldmath Reconstructions of Horndeski subclasses by gaussian processes}

\author[a,b]{Marco C. Mendoza-Marcos\orcidlink{0009-0006-6948-3778}}
\author[a,b]{and Ana A. Avilez-López\orcidlink{0000-0003-2223-4716}}
\affiliation[a]{Facultad de Ciencias F\'{\i}sico Matem\'aticas, Benem\'erita Universidad Aut\'onoma de Puebla,\\Apdo. Postal 1152, Puebla, Pue., M\'exico}
\affiliation[b]{Centro Internacional de F\'isica Fundamental, Benem\'erita Universidad Aut\'onoma de Puebla, Apdo. Postal 1152, Puebla, Pue., M\'exico.}

\emailAdd{marco.mendozamarcos@viep.com.mx}
\emailAdd{aavilez@fcfm.buap.mx}

\abstract{In this work we reconstruct two subclasses of theories from the Horndeski class, the most general scalar-tensor theories of gravity with second-order equations, according to a variety of late-universe cosmological data: model independent measurements of the Hubble parameter from Cosmic Chronometers, as well as Hubble parameter and distance measurements from Baryonic Acoustic Oscillations. In the modified gravity family considered, the scalar field is non-minimally coupled to gravity leading to time variations in the gravitational constant at cosmological scales. The theories considered are also consistent with constraints on the propagation speed of gravitational waves. The reconstruction is carried out by using the Gaussian Processes bayesian technique either for the Brans--Dicke and the Cubic Galileon subclasses of Horndeski theory. We find that the most likely Gaussian Process reconstruction for the data favors a cosmology from the Brans--Dicke subclass over $\Lambda$CDM in the range $z\in(0.4, 1.7)$, and for the Cubic Galileon subclass we find that the most likely reconstruction indeed favors modifications of the scalar field to gravity with evolving dark energy density in the whole range of redshift observations.}


\arxivnumber{2607.14154}

\begin{document}
\maketitle
\flushbottom

\newpage
\section{\label{sec:intro}Introduction}

So far General Relativity (GR) has been a successful theory describing gravity at scales ranging from millimeters to astronomical units with outstanding precision \cite{Will:2018}. However, it is not yet clear whether this theory provides a complete and consistent description of gravity at the largest scales. Particularly, a consistent explanation for the accelerated expansion of the universe is incomplete within the standard $\Lambda$CDM cosmological model based on Einstein's GR, where the cosmological constant $\Lambda$ accelerates the expanding universe. Actually, the cosmological constant problem addresses that it may not be constant but rather a time-varying dark energy (DE) density, and still the existence of profound open questions regarding the nature of DE demands that the possibility that gravity operates in a different way than the GR prescription must be tested.

Along the last decades, a host of metric theories describing  gravity as an effect of curvature of space-time, just as GR but according to a different variational principle, emerged by relaxing the different conditions imposed by Lovelock's theorem. In particular, when the condition that the metric $g_{\mu\nu}$ is the only field in the action, is relaxed one can consider other fields in the gravitational sector. A wide set of these modified theories of gravity are comprehended within the Horndeski scalar-tensor class. Within these metric theories, a scalar degree of freedom is added to the gravitational sector such that the field equations remain second order. In  this scenario the scalar might be responsible for the accelerated expansion of the Universe by means of kinematical and dynamical self-interaction terms in the action of the theory.

Scalar-tensor theories of gravity are well established and studied theories in the literature; they may arise as the compactification of higher dimensional theories such as Kaluza-Klein ~\cite{ModernKaluzaKlein} or the Dvali-Gabadaze-Porrati theory ~\cite{Dvali:2000hr}. They are also used as a simple way to model variations for the Newton gravitational constant by means of non-minimal coupling between the scalar field and the Ricci scalar, that on the largest scales corresponds to a time-varying gravitational coupling in contrast to the universal $G_N$ that also appears in the action for GR. But actually, can we think of the Newton constant to be the same even on the largest scales? We think the answer may be no, which motivates us to study the subclass of gravity models from the Horndeski theory on the cosmological scales with the so-called non-minimal coupling.

The possibility of a varying gravitational constant was picked up in the 1960s by Brans and Dicke, who developed the prototypical version of what are now considered scalar-tensor theories of gravity, by implementing Mach's principle to GR ~\cite{Clifton:2011jh, Brans-Dicke}. The Brans--Dicke theory is a realization within the Horndeski family and is one of the simplest extensions of GR depending on one additional parameter $\omega$. Solar System data put very strong constraints on the Brans--Dicke parameter whilst at the cosmological level there have been investigations constraining this parameter using CMB data and LSS measurements ~\cite{Avilez:2013}.

The Horndeski family is also known in the literature as the generalized covariant Galileon theory ~\cite{Kobayashi_2019}. The covariant Galileon has been investigated for its interesting cosmological solutions and its capability to accelerate the universe without the need of cosmological constant. Many investigations on the covariant Galileon have followed the form of the action in which only the minimal coupling of the scalar-field is considered. These models have been ruled out by observations ~\cite{Renk_2017}. However, Galileon models with non-minimal coupling admit regions of the parameter space that are in agreement with observational data. ~\cite{Zumalacarregui:2020}.

Although testing laws of gravity beyond the solar system scales, specifically at cosmological scales, is necessary to understand our universe and on itself is one of the most fascinating open problems in modern physics, the deep open questions behind the actual cosmological concordance paradigm such as the cosmological constant, the Hubble tension and our lack of knowledge about the nature of dark energy are undeniably good reasons not only for testing large classes of cosmological models in a systematic way against observations, but furthermore are signs of a clear lack of a complete and consistent cosmological theoretical framework to describe the universe. Nowadays, we urge for efficient methods for seeking for a meaningful and consistent cosmology. On the other hand, nowadays there is a large variety of theoretical proposals for achieving such improvement and hence it is convenient to sort them in large classes of models parametrized by free functions. These classes are to be constrained by observations by taking minimal theoretical assumptions into consideration. A bayesian approach to constrain these kind of classes, is to derive a set of posterior distributions for the free functions as random variables according to specific observational datasets. A convenient and natural technique to determine such distributions largely used in cosmology, which usually are called reconstructions of free functions, is that of Gaussian Processes \cite{Seikel_2012}.

In this work we carry out reconstructions of the Brans--Dicke and Cubic Galileon subclasses, both realizations of the Horndeski action, using the Gaussian Processes bayesian technique. We use data of Hubble parameter measurements coming from the direct model-independent method of Cosmic Chronometers (CC) \cite{Simon_2005} and points from BAO (SDSS-BOSS) \cite{Alam_2017, Vargas_Maga_a_2018} to increase the statistics, and other dataset of angular distance measurements from BAO (DESI DR2) \cite{Adame_2025_3, Abdul_Karim_2025} . We find for the case of the Brans--Dicke model that the parameter $\omega$ has a dependence on $z$ using  Hubble parameter data. The preference over this gravity model in the redshift range $z\in(0.4, 1.7)$ is addressed. In the case of the Cubic Galileon model, we find that the best-fit models for the Hubble parameter and for the angular diameter distance over the entire redshift range of the observations is always in this subclass, meaning the data favor modifications of gravity with the feature that the dark energy density evolves in time.

The outline of this paper is as follows, in section \ref{sec:horndeski} we introduce the general aspects of the Horndeski scalar-tensor class. In sections \ref{sec:reconstructions} and \ref{sec:teo_reconstructions}, the methodology for the reconstruction process is treated, here the Gaussian Processes for the datasets are derived, and the reconstructions of the Horndeski subclasses are performed. The discussion of the results is made in section \ref{sec:discussion}. In section \ref{sec:conclusions}, the conclusions of this work are summarized. The appendix section contains theoretical aspects of the Horndeski subclasses and of Gaussian Processes as well as a subsection on the priors, initial conditions and binning selection used in this work.

\section{\label{sec:horndeski}Scalar-tensor Horndeski class in the Jordan frame}
An standard procedure to modify gravity is to relax the assumptions imposed by Lovelock's theorem ~\cite{Lovelock:1971yv, Lovelock:1972vz} in GR. One possible modification is to assume that the theory of gravity depends not only on the rank-2 metric tensor $g_{\mu\nu}$ but on additional fields in the gravitational sector. The simplest scenario is to consider an extra scalar field $\phi$. 

The most general four-dimensional scalar-tensor theory of gravity with second-order field equations, generally covariant, Lorentz invariant, is the Horndeski class of theories described \newpage by the action ~\cite{Deffayet_2011, Horndeski:1974wa},

\begin{eqnarray}
    \label{HorndeskiAction}
    S=&&\frac{1}{16\pi}\int
    \left[G_{2}(\phi,X)-G_{3}(\phi,X)\Box\phi+G_{4}(\phi,X)R +G_{4X}\left[(\Box\phi)^2-\phi^{\mu\nu}\phi_{\mu\nu}\right]+G_{5}(\phi,X)G^{\mu\nu}\phi_{\mu\nu}\right. \nonumber \\
    &&\left.-\frac{G_{5X}}{6}\left[(\Box\phi)^3-3    \Box\phi\phi^{\mu\nu}\phi_{\mu\nu}+2\phi_{\mu\nu}\phi^{\nu\lambda}\phi^{\mu}_{\lambda}\right]\right]\sqrt{-g}d^4x+\int\mathcal{L}_m(\Psi,g_{\mu\nu})\sqrt{-g}d^4x,
\end{eqnarray}
where $R$ is the Ricci scalar, $X\equiv-\frac{1}{2}\nabla^\mu\phi\nabla_\mu\phi$, $\phi_\mu\equiv\nabla_\mu\phi$ and $\phi_{\mu\nu}\equiv\nabla_\mu\nabla_\nu\phi$. This is also the action for the so-called generalized covariant Galileon ~\cite{Deffayet_2009, Deffayet_2009_2}. We note that the lagrangian for the matter content depending on the matter fields $\Psi$ is coupled to the metric but not to the scalar field, this is because the conformal frame choice is the Jordan frame ~\cite{PhysRev.125.2163}, in which there is no interaction of matter fields with the scalar field so that point particles follow geodesics of the metric only, thus the Einstein equivalence principle is satisfied and the energy-momentum tensor of matter is conserved 
\begin{equation}
\label{energymomentumconsv}
\nabla_\mu T^{\mu\nu}=0.
\end{equation}

\subsection{\label{subsec: Vainshtein}Vainshtein screening mechanism}

In the scalar-tensor modified gravity theory described by the action (\ref{HorndeskiAction}) an extra force is mediated by the scalar degree of freedom. This force must be screened on small scales where GR has been well tested, i.e. in the lab or solar system. This is achieved by either making the scalar field effectively massive in the vicinity of a source, which is known as the chamaleon mechanism ~\cite{Khoury_2004_1, Khoury_2004} or by making the field effectively weakly coupled to the source which is the basics of the Vainsthein mechanism~\cite{Vainshtein:1972sx, Babichev_2013}. There are other screening mechanisms such as the symmetron \cite{Hinterbichler_2010} or k-Mouflage \cite{BABICHEV_2009}.

The Vainshtein mechanism achieves to suppress the effects of the scalar field for scales smaller than a characteristic scale known as the Vainshtein radius. In the case of a spherical distribution of non-relativistic matter with mass $\mathcal{M}$, the corresponding Vainshtein radius is

\begin{equation}
    r_V\coloneq\left(\frac{\mathcal{M}}{8\pi M_{Pl}\Lambda^3}\right)^{1/3},
\end{equation}
where $\Lambda$ defines the energy scale at where modifications of gravity become important. Assuming spherical symmetry the scalar field accounts for the present accelerating expansion of the universe if $\Lambda$ is of the order
\begin{equation}
    \label{Energyscale}
    \Lambda\sim\left(M_{Pl}H_0^2\right)^{1/3}.
\end{equation}
For a solar mass, this gives $r_V\sim100$pc, which is much larger than the size of solar system and well within the Hubble radius. This estimation of the energy scale $\Lambda$ will be relevant to impose priors on the functions we aim to reconstruct.  

\subsection{\label{subsec:gw170817}Horndesky subclass consistent with GW170817 constraints}
After the simultaneous detection of gravitational waves GW170817 and the $\gamma$-ray burst GRB170817A ~\cite{Abbott2016} providing a tight constraint on the propagation speed of gravitational waves $c_{GW}$, modified theories of gravity can be tested. The bounds on the difference between $c_{GW}$ with the speed of light in natural units imposed by this event are
\begin{equation}
    -3\times10^{-15}<c_{GW}-1<7\times10^{-16}.
\end{equation}
Therefore only a subclass of Horndeski theories should be able to predict $c_{GW}=1$. In effect, the prediction for $c_{GW}$ given by Horndeski theories is
\begin{equation}
    c_{GW}=\frac{G_4-X(\ddot{\phi}G_{5X}+G_{5\phi})}{G_4-2XG_{4X}-X(H\dot{\phi}G_{5X}-G_{5\phi})}.
\end{equation}
In order for this quotient to be equal to one, it must be satisfied:  
\begin{eqnarray}
    G_{4X}=0, \quad G_5=0.
\end{eqnarray}
Thus a subclass within Horndeski class compatible with this constraint is described by the action
\begin{eqnarray}
\label{galaction}
    S=&&\frac{1}{16\pi}\int
    \left[G_{2}(\phi,X)-G_{3}(\phi,X)\Box\phi+G_{4}(\phi)R\right]\sqrt{-g}d^4x+ S_m[\Psi, g_{\mu\nu}].
\end{eqnarray}
The corresponding general equations of motion derived from this action are presented in appendix \ref{app:a}.

It is important to mention that the modified gravity models of interest in this work which are described by the action (\ref{galaction}), might be only valid to describe late-time cosmic evolution, given that the physics of gravitational waves at earlier stages may be quite different, therefore the previous constraint may not apply since the detected GW so far come from local sources, however the latter case is out of the scope of this work. 

\subsection{\label{subsec:nonminimalcoup}Non-minimally coupled Horndeski subclass}
The term $G_4(\phi)R$ is known as the non-minimal coupling of the scalar field to the curvature, implying that the gravitational effects are due not only to the curvature of space-time but also to variations of the scalar field acting as a gravitational coupling to matter. In many theories corresponding to specific realizations of the Horndeski class such as the Brans--Dicke or the Cubic Galileon theories, $G_4(\phi) \rightarrow\phi$ is commonly interpreted as such dynamical gravitational coupling.

The importance of the non-minimal coupling relies on that, if only the minimal coupling is considered, i.e. $G_4(\phi)$ is constant (usually $\frac{M_P^2}{2}$), then the theory predicts the growth of metric potentials ~\cite{Barreira_2014} which is in tension with observational evidence of the Integrated Sachs-Wolf (ISW) effect ~\cite{Pl2016, Ferraro_2015} that exhibits a positive correlation with CMB temperature anisotropies ~\cite{Ho_2008} that is due to the measure of decaying metric potentials that are well predicted by $\Lambda$CDM. However, when the non-minimal coupling is considered, the model is able to predict decaying potentials in some regions of the parameter space. Another important reason is the recent cosmological evidence for the non-minimal coupling. In light of new data released by DESI, we can see a preference of the data for dark energy with a time-varying equation of state given by the CPL parametrization ~\cite{CHEVALLIER_2001} over $\Lambda$CDM. Nevertheless, recently in ~\cite{Wolf_2025} we learned that a model of dark energy described as a non-minimally coupled scalar-field is preferred over CPL in terms of a Bayes factor using a BAO+CMB data combination. With these theoretical and phenomenological reasons we stress that the non-minimal coupling is important to keep as a consistency requirement. 

\subsection{Cubic-galileon-like subclass}

In this work, we consider the simplest theories within the Horndeski class consistent with GW observations described by action (\ref{galaction}): the theory for the non-minimally coupled Brans--Dicke free scalar and its kinetic-self-interacting version usually dubbed the Cubic Galileon. We refer to each case simply as a subclass.
In either subclasses, we associate the scalar to a time varying gravitational coupling as originally proposed by Brans and Dicke in order to account for the Mach principle. Modifications to gravity arise due to the dynamics of the scalar. The action describing these subclasses is:
 
\begin{eqnarray}
\label{cubicgal}
    S=\frac{1}{16\pi}\int\left[\phi R-\frac{\omega}{\phi}\nabla^{\mu}\phi\nabla_{\mu}\phi+\frac{\alpha}{8\phi^3}\nabla^\mu\phi\nabla_\mu\phi\Box\phi\right]\sqrt{-g}d^4x+S_m[\Psi, g_{\mu\nu}].
\end{eqnarray}
\vspace{2mm}

\noindent When the scalar is self-interacting $(\alpha\neq 0)$ it undergoes through the so called Vainsthein screening mechanism that suppresses its observational effects at scales where GR has been well tested ~\cite{Avilez-Lopez:2015}. Here $\phi$ is given in units of $M_{Pl}^2$, $\omega$ is a dimensionless parameter controlling the purely scalar-field dependent kinetic term; $\alpha$ is also a dimensionless parameter mediating the non-linear kinetic interactions that effectively couple derivatives of the Galileon and the metric through a process commonly dubbed as ``kinetic gravity braiding'' \cite{Pujol_s_2011, C_dric_Deffayet_2010}. At perturbations level, this process modifies the way in which gravitational potentials respond to density fluctuations in comparison to the $\Lambda$CDM model. 

It is well known that for spherically symmetric sources, the dimensionless parameter $\alpha$ is related to an energy scale $\Lambda$ as follows 

\vspace{1mm}
\begin{equation}
\label{alpha_funlambda}
\alpha = \left(\frac{M_{Pl}}{\Lambda}\right)^3.
\end{equation}
\vspace{2mm}

\noindent If the energy scale is of the order (\ref{Energyscale}), then the scalar field dynamics is able to give rise to the accelerated expansion of the universe,  therefore, in our analysis we take $\alpha$ values $\sim M_{Pl}^2H_0^{-2}$ in order that modifications to gravity arise at cosmic scales.

\vspace{3mm}

Equations of motion for this theory result from specifying $G_2(\phi, X)=\frac{2\omega}{\phi}X$, $G_3(\phi, X)=\frac{\alpha}{4\phi^3}X$ and $G_4(\phi)=\phi$ in (\ref{tensoreqGal}) and (\ref{scalareqGal}). As we are interested in studying the expansion of the universe we consider the background cosmological solutions, therefore we reduce these equations for a spatially-flat FLRW space described by the following line element 
\vspace{3mm}
\begin{equation}
    ds^2=-dt^2+a^2(t)\left\lbrace dr^2+r^2(d\theta^2+\sin^2{\theta} d\phi^2)\right\rbrace.
\end{equation}
\vspace{2mm}

 Two independent equations result corresponding to time and spatial components of the tensor equation (\ref{tensoreqGal}), respectively
\begin{eqnarray}
\label{firstgaleq}
    H^2=\frac{8\pi\rho}{3\phi}-H\frac{\dot{\phi}}{\phi}+\frac{\omega}{6\phi^2}\dot{\phi}^2+\frac{\alpha}{8\phi}\left[H\left(\frac{\dot{\phi}}{\phi}\right)^3+\left(\frac{\dot{\phi}}{\phi}\right)^4\right],
\end{eqnarray}

\begin{eqnarray}
\label{secondgaleq}
    \Box\phi=8\pi P\phi+3H^2\phi+2\dot{H}\phi+\frac{\omega}{2\phi}\dot{\phi}^2+\frac{\alpha}{8\phi^3}\left(\ddot{\phi}-\frac{3\dot{\phi}^2}{2\phi}\right)\dot{\phi}^2,
\end{eqnarray}
and the scalar field equation from (\ref{scalareqGal})
\begin{eqnarray}
\label{thirdgaleq}
    8\pi(-\rho+3P)=(2\omega+3)\Box\phi+\frac{3\alpha}{4\phi^2}\left(3H^2\dot{\phi}^2+2H\dot{\phi}\ddot{\phi}+\dot{H}\dot{\phi}^2+\frac{3\dot{\phi}^2\ddot{\phi}}{2\phi}-\frac{5\dot{\phi}^4}{2\phi^2}\right).
\end{eqnarray}
This is a highly non-linear second-order system of ordinary differential equations, as was expected from the modificated-- Lovelock's theorem. Equations (\ref{secondgaleq}) and (\ref{thirdgaleq}) determine the dynamics of the metric and the scalar, whilst (\ref{firstgaleq}) acts as a constriction equation.

As mentioned above, for pedagogical reasons, we carry out a separate analysis for the simplest case without self-interactions ($\alpha=0$) corresponding to the Brans--Dicke-like subclass which is a well established theory of gravity known to be indistinguishable from GR when the dimensionless parameter $\omega\rightarrow\infty$. 
As mentioned before, cosmological constraints on the Brans--Dicke parameter $\omega$ have been derived using CMB and LSS data in \cite{Avilez:2013}, the obtained lower limit $\omega>890$ gives rise to a derived constraint for the cosmological effective gravitational constant relative to the solar system measurement given by  $0.981<G_{eff}/G_N<1.285$. In the following, we consider that $\omega\sim \mathcal{O}(10^0-10^4)$ in order to study the cosmic expansion alone for a wide range of realizations of the Brans--Dicke subclass, including either those similar to GR and those with significant variations of the gravitational coupling for small values of $\omega$. 

Regarding the matter content of the universe, we consider a mixed fluid composed by matter and dark energy species described as perfect fluids with barotropic equation of state $P_I=w_I \rho_I$, where the constant $w_I$ is the parameter of state and the index $I=m,\Lambda$ stand for dust and dark energy respectively. Since we use the Jordan frame, species are thermally decoupled and then the standard continuity equation describes the evolution of the energy density of each species according to (\ref{energymomentumconsv}), whose solution as function of redshift $z$ is
\begin{equation}
\label{energydensity}
    \rho_I = \rho_{0I}(1+z)^{3(1+w_I)},
\end{equation}
here $\rho_{0I}$ is the energy density of each species today. It is well known that, in contrast to the Cubic Galileon, the Brans--Dicke class is not able to reproduce the accelerated expansion without considering a dark energy component. For sake of simplicity, for that subclass, we consider the cosmological constant ($w_\Lambda=-1$) in order to achieve such prediction. We also consider a cosmological constant within the Cubic Galileon subclass but setting free the energy density parameter, in order to figure out in what extent this kind of fluid is needed to reproduce the cosmic accelerated expansion.

\begin{figure}
\centering
\includegraphics[scale = 0.6]{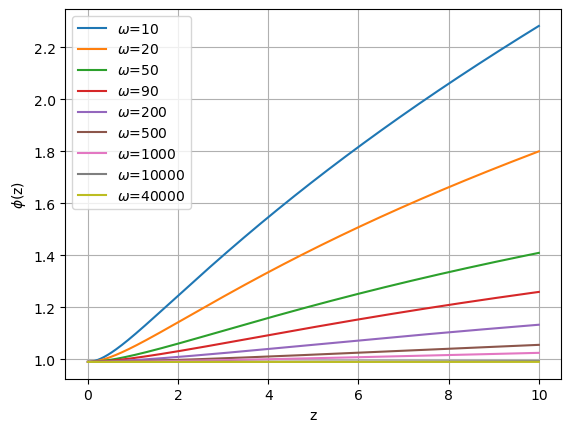}
\caption{\label{fig:BDscalarfields} Solutions of the scalar field $\phi(z)$ in the Brans--Dicke subclass for different values of the dimensionless parameter $\omega$. The $y$-axis has units of $M_{Pl}^2$. For the value $\omega\sim10^4$, the field doesn't evolve, as expected from the GR limit of the Brans--Dicke theory.}
\end{figure}

In figure \ref{fig:BDscalarfields} solutions of the scalar field in the Brans--Dicke subclass for different values of $\omega$ are shown, the units in the y-axis are $M_{Pl}^2$. Notice that for $\omega\sim10^4$ the field practically remains constant and equal to $M_{Pl}^2$, so GR is recovered.

\begin{figure}
\centering
\includegraphics[scale = 0.6]{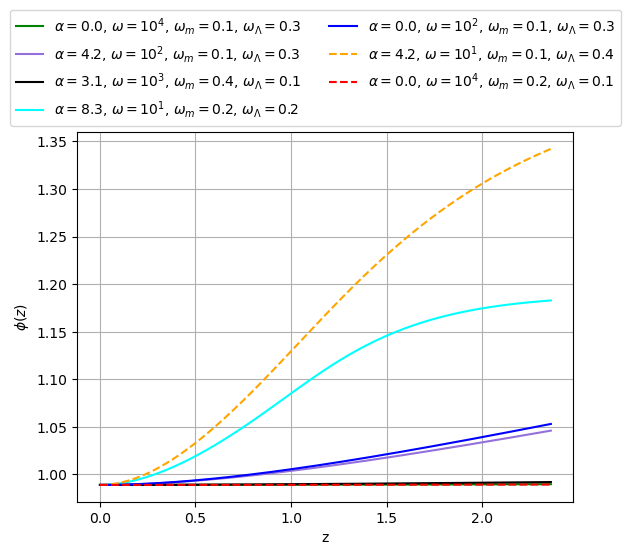}
\caption{\label{fig:phi_CG} Solutions of the scalar field $\phi(z)$ for different combinations of parameters in the Cubic Galileon subclass. The evolution of the scalar field depends strongly on the combination of the four parameters. The $y$-axis has units of $M_{Pl}^2$.}
\end{figure}

\begin{figure}
\centering
\includegraphics[scale = 0.6]{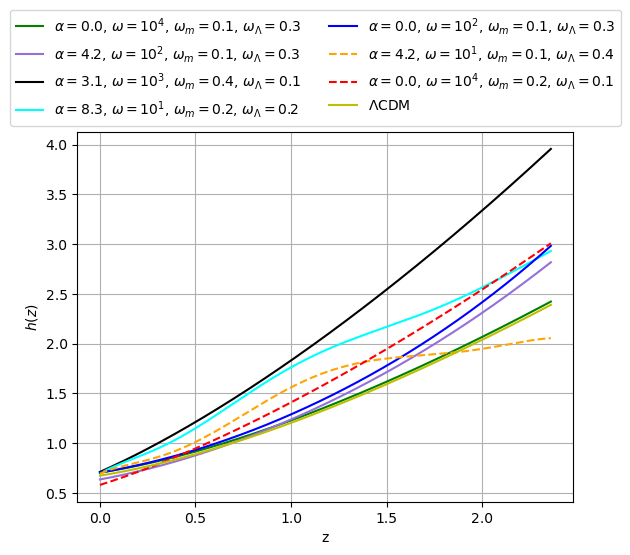}
\caption{\label{fig:h_CG} Solutions of the Hubble parameter $h(z)$ for different combinations of parameters in the Cubic Galileon model.}
\end{figure}

Figures \ref{fig:phi_CG} and \ref{fig:h_CG} show solutions for the scalar field $\phi(z)$ and the normalized Hubble parameter $h(z)= H(z)/100km/s/Mpc$, for different combinations of parameters in the Cubic Galileon subclass, $\alpha$ and $\omega$, and the physical energy density parameters today, $\omega_{0m}$ and $\omega_{0\Lambda}$. Notice that the evolution of the scalar field depends on the combination of these four parameters as well as the Hubble parameter $h(z)$. The initial conditions $\phi_0$ and $\phi_0^\prime$ were set according to different criteria discussed in appendix \ref{app:e}.

It is well known that the Cubic Galileon is able to account for the accelerated expansion of the universe, this can be seen in figure (\ref{fig:h_CG}): the $\Lambda$CDM prediction for the expansion rate (yellow line) is practically reproduced by a particular realization within the Cubic Galileon subclass (green line) with a lesser amount of dark energy.

\section{\label{sec:reconstructions}Model-independent reconstructions by gaussian processes}
In this section, we discuss the methodology for obtaining model-independent reconstructions using Gaussian Processes (GP) of different observational datasets to evaluate the posterior distribution in a parameter space of a given theory. A review of the Gaussian Processes formalism is presented in appendix \ref{app:b}.

\subsection{The data}

Since in this work we are interested in studying the accelerated expansion of the universe within the Horndeski subclasses mentioned above, the corresponding reconstructions are inferred from distance measurements of different distant objects and other observables characterizing the background expansion of the universe. For that purpose, we consider two datasets of this kind. Firstly, Hubble parameter measurements at different redshifts from Cosmic Chronometers (CC) which are independent of a cosmology since they are based on the quantity $dz/dt$ which is inferred by observing two ensembles of passively-evolving galaxies at different redshifts with the only assumption of the modeling of stellar ages through stellar population synthesis techniques \cite{Gomez-Valent_2018}, therefore the quantity 
\begin{equation}
  H(z)=(1+z)^{-1}\frac{dz}{dt}  
\end{equation}
can be measured. The corresponding data is reported in different works such as  ~\cite{Gomez-Valent_2018, Vagnozzi_2021, Vel_zquez_2024, Gadbail_2024}, with the addition of Hubble parameter measurements from BAOs by BOSS-SDSS to increase the statistics.

Secondly, we use a recently realeased dataset of the angular diameter distance measurements of BAO from DESI-DR2 \cite{Adame_2025_4}. Although, these measurements are model-dependent (as we shall see in equations (\ref{AlPac})), we include them in order to consider diverse observables characterizing the same physical process -namely the expansion-. Indeed, the distance measurements of BAO provided by different surveys are usually not directly determined, the actual observable is the ratio of distance and $r_d$ given by:

\begin{equation}
r_d=\int_{z_d}^{\infty}\frac{c_s(z)}{H(z)}dz.
\end{equation}
Thus BAO measurements rely on the Alcock-Paczynski-like dilation parameters which correspond to longitudinal and transverse distances to the line of sight measurements quantified by 

\begin{equation} \label{AlPac}\alpha_{\parallel}=\frac{D_H(z)r_d^{fid}}{D_H^{fid}(z)r_d},\quad \quad \alpha_{\bot}=\frac{D_M(z)r_d^{fid}}{D_M^{fid}(z)r_d},
\end{equation}
\vspace{2mm}

\noindent where the quantities with label \textit{fid} are given by a fiducial $\Lambda$CDM model, $D_M(z)$ is the angular diameter distance and $D_H(z)$ the Hubble distance. By using a fiducial cosmology as a reference, the constrained quantities reported by DESI are $D_M(z)/r_d$ and $D_H(z)/r_d$. In this work we analyze the former dataset consisting of six points at redshifts $z\in\lbrace0.510, 0.706, 0.934$ $1.321, 1.484, 2.330\rbrace$ traced by galaxies, quasars and the Lyman-$\alpha$ forest.

\subsection{Gaussian process reconstructions of observables}
In this section we report the Gaussian Processes reconstructions corresponding to each dataset described above. Once these model-independent reconstructions in the redshift-observable plane are derived, they are used to construct a posterior distribution function in the space of parameters of the theoretical model, reported in a further section \ref{sec:teo_reconstructions}. It is worth to mention, that an outstanding advantage of this technique is that the posterior can be computed without evaluating a likelihood function in the parameter space, which usually is highly numerically expensive. 

\subsubsection{Gaussian process reconstruction of Hubble parameter data}
Following the methodology explained in appendix B, a Gaussian Process was constructed corresponding to the normalized Hubble parameter $h(z)$, for different redshifts running within  $z\in(0,2.3)$. In order to improve the numerical performance and the quality of the reconstruction the squared exponential kernel (\ref{sqexp}) was used. The resulting optimized hyperparameters are $\sigma_f = 1.64$ and $l=2.43$. 

\begin{figure}
\centering
\includegraphics[scale = 0.4]{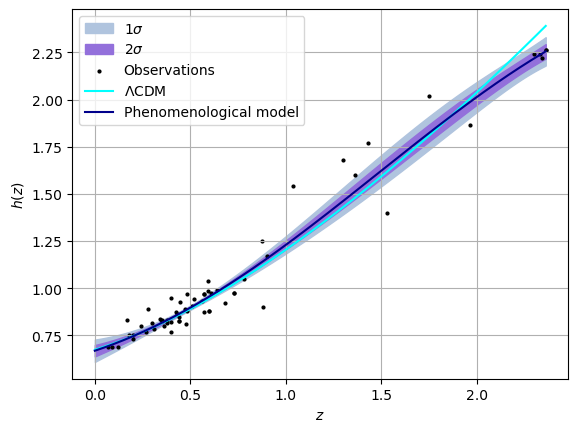}
\caption{\label{fig:gpforh} Gaussian Process for the Hubble parameter data from CC and BAO-BOSS-SDSS and observation points. The $\Lambda$CDM is shown for comparison. Significant differences between $\Lambda$CDM and the GP predictive mean happen at $z>2$.}
\end{figure}

In figure \ref{fig:gpforh}, the predicted mean values and their $1\sigma$ and $2\sigma$ limits for the redshift domain are shown, along with the concordance $\Lambda$CDM model Planck-2018 best-fit. Notice that the $\Lambda$CDM model is not always the most likely cosmology according to this reconstruction, the most significant deviations occur at $z\gtrsim2$. Actually, the continuous set of mean values of this Gaussian Process reconstruction corresponds to a phenomenological model fully independent of theoretical considerations. In order to roughly test the goodness-of-fit to the data of this phenomenological and the $\Lambda$CDM models, we evaluate the $\chi^2$ respectively, and the results are shown in table \ref{tab:chisq_h}. According to this test, the Gaussian Process phenomenological model fits better the Hubble parameter data than the $\Lambda$CDM model.  

\begin{table}
    \centering
        \begin{tabular}{|c|c|}
        \hline
        \textbf{Model} & $\mathbf{\chi^2}$ \\
        \hline
        GP phenomenological  & 30.77\\
        $\Lambda$CDM & 43.65 \\
        \hline
        \end{tabular}
    \caption{Results of the $\chi^2$ test on the GP reconstruction for the Hubble parameter data and the $\Lambda$CDM model. The GP phenomenological model shows to be a better fit for the observations by this test.}
    \label{tab:chisq_h}
\end{table}

\subsubsection{Gaussian process reconstruction of BAO angular diameter distance data}
\begin{figure}
\centering
\includegraphics[scale = 0.4]{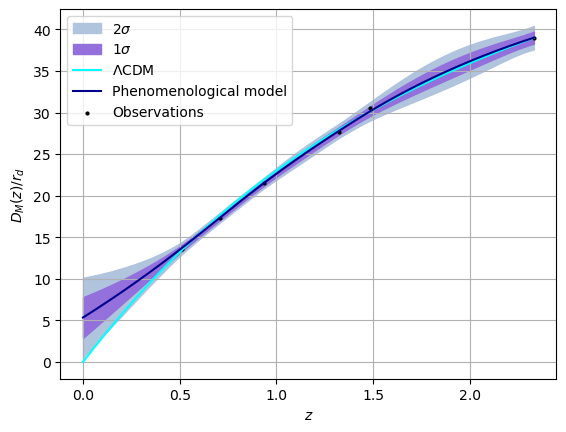}
\caption{\label{fig:gpforDM} Gaussian Process for the BAO-DESI-DR2 angular diameter distance distance data and observation points. The $\Lambda$CDM model prediction is plotted for comparison. Significant deviations between the two models are found in $0<z<0.5$ due to the lack of observations at these redshift values.}
\end{figure}

The Gaussian Process reconstruction corresponding to the angular diameter distance of BAOs data from DESI-DR2 is shown in figure \ref{fig:gpforDM}. The kernel used in this reconstruction was the Matern kernel (\ref{matern}), and the resulting hyperparameters are $\sigma_f=43.6$, $l=7.81$ and $\nu=1.5$. The solid lines correspond to the predicted mean values of the Gaussian Process phenomenological reconstruction for this dataset. Shaded bands correspond to $1\sigma$ and $2\sigma$ regions respectively. The concordance $\Lambda$CDM model exhibits differences from the phenomenological one within $0<z<0.5$ as expected, given the observation points are scarce at these redshift values. 

\begin{figure}
\centering
\includegraphics[scale = 0.42]{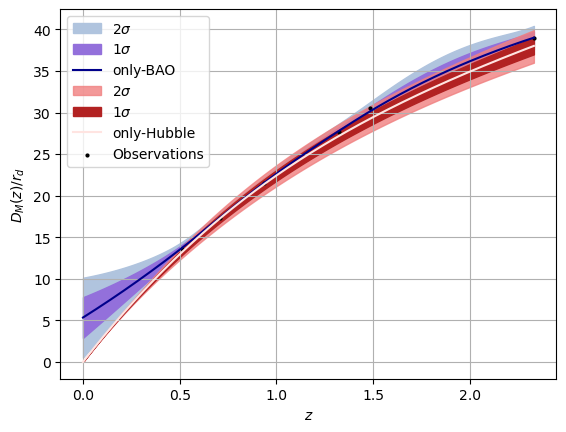}
\caption{\label{fig:priorDM} Individual GP for angular diameter distance from BAO-DESI-DR2 and Hubble parameter data.}
\end{figure}

\begin{figure}
\centering
\includegraphics[scale = 0.42]{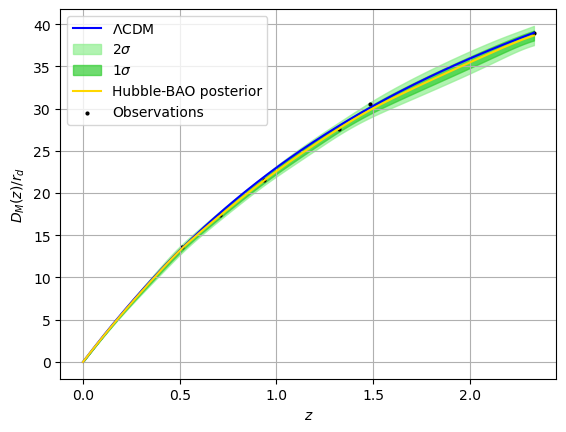}
\caption{\label{fig:postDM} Combined posterior GP from BAO-DESI-DR2 and Hubble parameter data. The concordance $\Lambda$CDM model is shown for comparison.}
\end{figure}

In order to carry out the combined bayesian analysis according to Hubble parameter and BAO observations, we assume that both reconstructions derived previously,  are independent distributions since individual datasets are uncorrelated. An important advantage of this combined analysis is that including observations over all the redshift range helps to overcome the lack of BAO observations at $0<z<0.5$. Therefore through a simple product we combine the reconstruction for the only-$h(z)$ GP with the only-BAO GP.  The resulting $h(z)$-BAO Gaussian Process reconstruction for the combined datasets is shown in figure \ref{fig:postDM} and in \ref{fig:priorDM}, the Gaussian Processes phenomenological models for individual datasets.

Likewise, as we did above for the $h(z)$ reconstruction,  we perform a rough $\chi^2$ assesment of the goodness of fit of different models to $D_M/r_d$ data: the only-BAO GP, only-$h(z)$ GP, $h(z)$-BAO GP and the $\Lambda$CDM model. The results are presented in table \ref{tab:chisq_bao}. 

As expected, given the small number of points within the DESI-BAO sample, the corresponding phenomenological model has the smallest $\chi^2$. In contrast, the combination of $h(z)$-BAO data provides a fit with similar and slightly larger $\chi^2$ which is statistically more robust since the sample is larger. Interestingly, both phenomenological models fit better the combined data than $\Lambda$CDM. The only-$h(z)$ phenomenological model yields the poorest fit, which suggests that fitting the BAO data requires a model beyond $h(z)$, implying different cosmic dynamics. 

\begin{table}
    \centering
    \begin{tabular}{|c|c|}
    \hline
        \textbf{Model} & $\mathbf{\chi^2}$ \\
        \hline
        only-BAO phenomenological & 1.12\\
        only-$h(z)$ phenomenolgical & 16.89\\
        $h(z)$-BAO phenomenolgical & 3.14\\
        $\Lambda$CDM & 8.55\\
        \hline
        \end{tabular}
    \caption{Results of the $\chi^2$ goodness-of-fit test on the different GP phenomenological models and the $\Lambda$CDM model. It can be noticed that the only-BAO GP phenomenological model fits better the observations according to this test, the second best-fit being the $h(z)$-BAO GP.}
    \label{tab:chisq_bao}
\end{table}

\section{\label{sec:teo_reconstructions}Reconstruction of the posterior distribution in the parameter space of the Horndeski subclasses.}

 The main purpose in this section is to reconstruct $\omega$ and $\alpha$ from (\ref{cubicgal}) at different redshifts using the Gaussian Process reconstructions corresponding to observations of Hubble parameter and distance measurements of BAO. However, at this point is important to explain the proposed idea for reconstructing the free ``parameters'' within the Horndeski subclasses. Firstly, let us consider a large sample of realizations of a given subclass, each of them corresponding to a set of parameters $\vec{\theta}=\{p_i\}_{i=1,..,N}$, that is, $p_1=\omega$ and $p_2=\alpha$ for the Cubic Galileon subclass. Secondly, consider a redshift partition corresponding to the Gaussian Process phenomenological reconstruction domain. Given a specific dataset $D$, for each value of $z$ within the partition, we can assign a weight or a probability to each element in the sample -corresponding to a realization of the theory- by using the corresponding Gaussian Process reconstruction, namely: 

 \begin{equation}\label{eq:weight}
\mathcal{P}(\vec{\theta},z|D)\propto \exp{\left\lbrace-\frac{(O_{\mathrm{pred}}(z,\{p_i\})-\bar{O}_{GP} (z))^2}{2(\sigma_{GP}(z))^2}\right\rbrace},
\end{equation}
\vspace{2mm}

\noindent where $O_{\mathrm{pred}}(z,\{p_i\})$ is the predicted observable by the theory, $\bar{O}_{GP}(z)$ and $\sigma_{GP}(z)$ are the mean  and  variance of the observable random variable Gaussian Process reconstruction corresponding to the dataset D, respectively. 

In this way, a posterior distribution in parameter space is constructed for each $z$ in the redshift domain. Given the high non-linearity of the mapping between a given realization of the model (which by construction is a random variable with normal distribution) and its prediction, the resulting distribution in parameter space is not a gaussian. Reconstructions of individual parameters $p_i$ correspond to marginalized posterior distributions given by:   

\begin{equation}
     \label{marginalization}   P(z,p_i)=\int P(z,p_i,p_j) d^{N-1}p_j, \qquad i\neq j.
    \end{equation}

\subsection{Reconstruction of the Brans--Dicke subclass}

Since the Brans--Dicke is the simplest case, it serves as an illustration of the procedure we aim to implement, hence we determine a reconstruction of $\omega$ using only the model independent $h(z)$ Gaussian Process determined in (\ref{sec:reconstructions}).

Firstly, the system of equations (\ref{firstgaleq}, \ref{secondgaleq}, \ref{thirdgaleq}) is solved numerically for $\alpha=0$ and using $z$ as independent variable instead of $t$. In order to solve the system we implemented the Runge Kutta 4 method in our own code developed in Python. The normalized Hubble parameter $h(z)\equiv H(z)/(100\,\mathrm{km/s/Mpc})$ corresponds to a dynamical variable, the total energy density $\rho$ and the pressure $P$ functions are expressed in terms of the physical density parameters for different species, that is, $\omega_{0m}\equiv\Omega_{0m}h_0^2$ and $\omega_{0\Lambda}\equiv\Omega_{0\Lambda}h_0^2$, where $\Omega_{0i}\equiv\rho/\rho_c$ are the cosmological density parameters and $\rho_c\equiv3H_0^2/8\pi$ . The system of dynamical equations are presented in appendix \ref{app:c}.

We fix the values for the physical density parameters to the Planck estimates ~\cite{2020}.  As mentioned above, within the Brans--Dicke subclass the cosmological constant density value is taken just as in Planck-$\Lambda$CDM in order to predict the accelerated expansion of the universe. Since the system of equations is first-order, the relevant initial conditions are fixed as follows: $\phi_0=G_{eff}/G_N$ is set to the value $\phi_0=0,989$ according to the constraints on the gravitational coupling mentioned before. Besides, the initial condition of $h_0$ is set to Planck best-fit $h_0=0,676$. Solutions are obtained for different values of $\omega$. Numerically, the value for $\omega$ where the BD model reproduces GR predictions is $\omega\sim10^4$ in accordance with constraints from solar system experiments. 

\begin{figure}
\centering
\includegraphics[scale = 0.42]{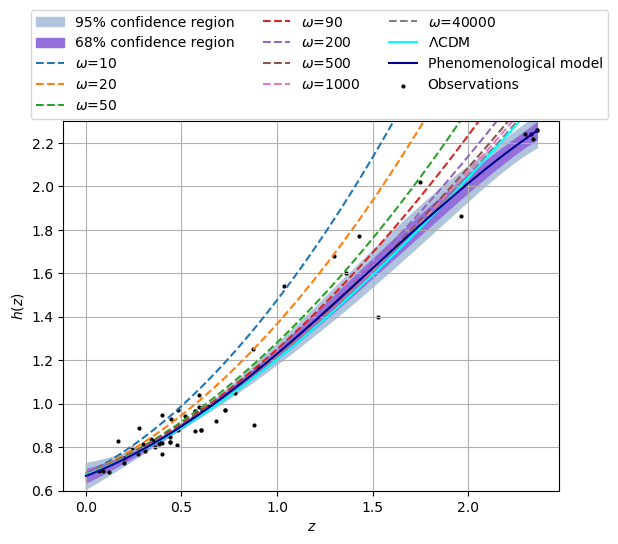}
\caption{\label{fig:BDandGP} A sample of solutions for the Hubble parameter $h(z)$ within the Brans--Dicke model for different values of the dimensionless constant $\omega$.}
\end{figure}

\vspace{3mm}

 Figure \ref{fig:BDandGP} shows solutions of $h(z)$ within the Brans--Dicke subclass, along with the GP reconstruction of the Hubble parameter at different redshifts. It can be noticed that the models tend to depart from the confidence region as $\omega$ decreases. Its also worth of stressing that some models are consistent with the model-independent $h(z)$ only for a short range of $z$ and becomes inconsistent at large redshifts. The $\Lambda$CDM model and the observations are also shown, notice that, likewise $\Lambda$CDM, the constant-$\omega$ Brans--Dicke subclass is unable to reproduce the most likely $h(z)$. The previous suggests that a subclass with a varying $\omega(z)$ would be better suited to reproduce observations.

\vspace{3mm}

In order to reconstruct a distribution parametrized along the redshift domain for $\omega$ as a random variable, we follow the procedure explained at the beginning of this section. Namely, the Gaussian Process reconstruction for an observable provides a distribution at every $z$ is used to assign a weight corresponding to the posterior probability to every Brans--Dicke model prediction of $h(z)$, especifically:
\begin{equation}
\mathcal{P}(\omega,z|D)\propto \exp{\left\lbrace-\frac{(h_{pred}(\omega,z)-\bar{h} (z))^2}{2(\sigma(z))^2}\right\rbrace},
\end{equation}
where $h_{teo}(z,\omega)$ is the model prediction, $\bar{h}(z)$ and $\sigma(z)$ are the mean and  variance of the model-independent Gaussian Process respectively. Note that this mapping depends on the redshift.

\vspace{3mm}

\begin{figure}
\centering
\includegraphics[scale = 0.8]{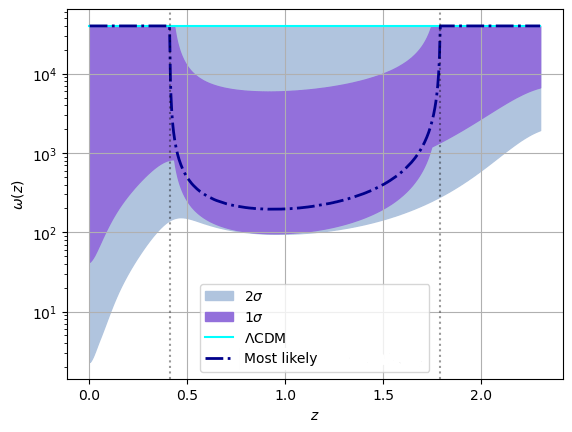}
\caption{\label{fig:BDrec} Reconstruction of the Brans--Dicke subclass dimensionless parameter $\omega$. Interestingly, within $0.4<z<1.7$, the Brans--Dicke cosmology is more likely than $\Lambda$CDM.}
\end{figure}

In order to evaluate the posterior $P(z,\omega|D)$, we swept the following region in parameter space:   $\omega\in[0, 40000]$. It is worth to mention that for implementing this method it is not necessary to evaluate a likelihood distribution, instead we use a model independent distribution for observations as in other likelihood-free bayesian methods \cite{Drovandi}. Figure \ref{fig:BDrec} shows the resulting posterior distribution, the most likely set of $\omega$ values at different redshifts and the $68\%$ and $95\%$ confidence regions. 

\vspace{3mm}

In summary, in order to determine a  \textit{reconstruction} for a free parameter of the BD subclass within a redshift domain, we firstly fit the data by using a Gaussian Process which provides a $z$-parametrized distribution. Besides, this distributions provides a non-parametric (phenomenological) model that best fits the data. Secondly, by sweeping an acceptable region of parameter space of the Brans--Dicke subclass, we compute the predictions corresponding to each point in that region (or realization of the subclass). Finally, we directly obtain the posterior distribution at each $z$ by evaluating the theoretical predictions in the Gaussian Process reconstruction of $h(z)$. In other words, we ``weight'' points in parameter space at each redshift using the model-independent Gaussian Process reconstruction (which can be thought as a ``distribution-like'' interpolation of data). Given that the Brans--Dicke subclass holds a single free parameter, the posterior distribution is actually the desired reconstruction.  

A compelling byproduct from this posterior distribution is the set of most-likely values of $\omega$ along the redshift range. Although the Brans--Dicke subclass studied here is a priori taken as a constant-parameters-deterministic framework, the a posteriori resulting $\omega(z)$ for an ensamble of deterministic-Brans--Dicke realizations can be used as a stochastic free-function reconstruction of a varying $\omega$ instead.  

Reconstructing this one-parameter subclass has been useful to illustrate the Gaussian Process free-function reconstruction method. Nevertheless, as in $\Lambda$CDM, within this subclass a cosmological constant is needed to account for the accelerated expansion of the universe and hence the cosmological constant problem and other open questions regarding the fundamental nature of this component remain up in the air. In this sense, a more appealing theory that is able to account for accelerated expansion with a lesser extent of a cosmological constant is the Cubic Galileon subclass, which has more than one parameter. In the next subsection, we carry out reconstructions for the  Cubic Galileon subclass by applying a more general procedure suitable for models with multidimensional parameter space.

\subsection{Reconstruction of the cubic galileon subclass}

The equations to be solved within this subclass are presented in appendix \ref{app:d}.  After algebraic operations in (\ref{galzeq2}) and (\ref{galzeq3}), we obtained a system of two equations for $\phi(z)$ and $\phi^\prime(z)$ and the constraint equation for $h(z)$, $h^\prime$ can also be expressed as a function of $\phi$ and $\phi^\prime$ and called wherever it appears in the equations.

 \subsubsection{Reconstruction using the $h(z)$ Gaussian Process}

In order to reconstruct the Cubic Galileon subclass, firstly, we solved the equations of motion in order to compute the theoretical prediction of the hubble rate $h_{pred}(\Vec{\theta}, z)$, for different realizations of $\Vec{\theta}=(\alpha, \omega, \omega_{0m}, \omega_{0\Lambda})$. These realizations form an ensemble of possible configurations predicted by model, and using the $h(z)$ model-independent Gaussian Process reconstruction we assign a probability to each ensemble element given by (\ref{eq:weight}), where the observable $O(z)$ in this case corresponds to $h(z)$. As explained in \ref{sec:teo_reconstructions}, by means of such mapping, the posterior distribution $P(\Vec{\theta},z)$ is evaluated . Afterwards, the marginalized posterior distribution is computed for individual parameters using (\ref{marginalization}).

\begin{figure}
\centering
\includegraphics[scale = 0.6]{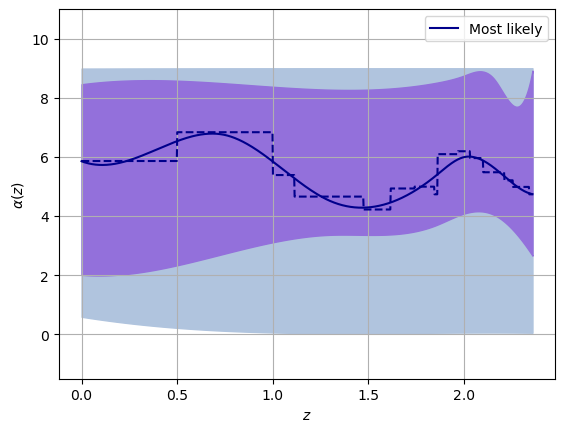}
\caption{\label{fig:alpha_h} Reconstruction of the self-interaction coupling  $\alpha$ according to the model-independent $h(z)$ GP in a four-dimensional parameter space. The reconstructed values of $\alpha$ correspond to a Vainshtein radius on cosmological scales.}
\end{figure}

\begin{figure}
\centering
\includegraphics[scale = 0.6]{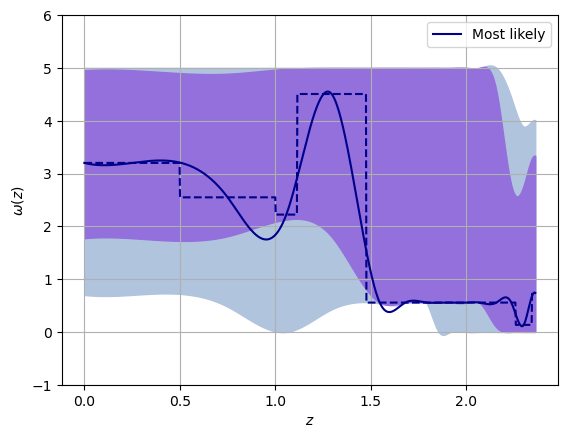}
\caption{\label{fig:omega_h} Reconstruction of  $\omega$ from the model-independent $h(z)$ GP in a four-dimensional parameter space. Since this parameter controls the strength of the modifications to gravity, the largest deviations from General Relativity occur around $z\gtrsim1.5$.}
\end{figure}

\begin{figure}
\centering
\includegraphics[scale = 0.6]{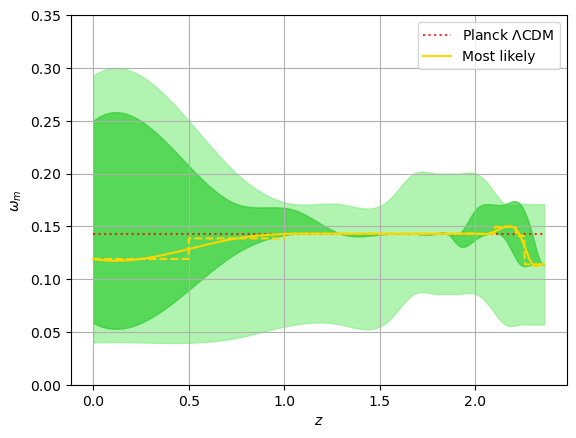}
\caption{\label{fig:omegam_h} Reconstruction of the physical density parameter $\omega_{0m}$ using the model-independent $h(z)$ GP in a four-dimensional parameter space. The Planck-$\Lambda$CDM best-fit is shown as comparison. The most likely value stays almost constant along the redshift domain and very close to Planck-$\Lambda$CDM.}
\end{figure}

\begin{figure}
\centering
\includegraphics[scale = 0.6]{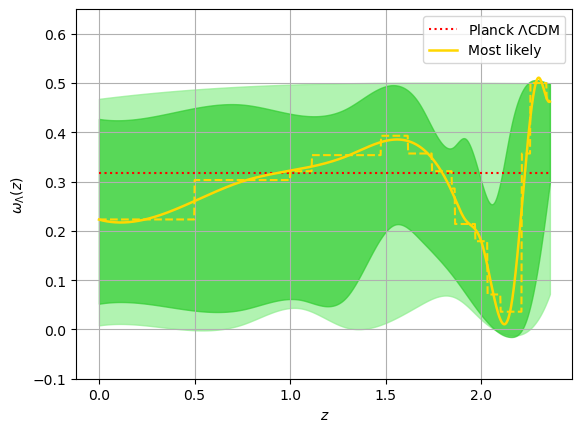}
\caption{\label{fig:omegal_h} Reconstruction of the physical density parameter $\omega_{0\Lambda}$ using the model-independent $h(z)$ GP in a four-dimensional parameter space. The Planck-$\Lambda$CDM best-fit is shown as comparison. The most likely value is changing over the bins and reaches two local maxima  in $z\sim 1.5$ and $z\sim 2.5 $, the minimum in $z\sim2$ is zero within the $69\%$ confidence level.}
\end{figure}

Figures \ref{fig:alpha_h}, \ref{fig:omega_h}, \ref{fig:omegam_h} and \ref{fig:omegal_h} show the resulting distributions which correspond to the so-called reconstructions of each parameter. The uncertainty of such reconstructions are encoded by means of the $68\%$(dark purple) and $95\%$(light purple) confidence regions while the most likely values corresponds to the dashed dark line. It is noteworthy, that in any case, the probability maxima tend to a constant along different redshift bins that arise naturally. This suggests that constant theoretical predictions are consistent with the observations within these emergent redshift bins, whereas outside them variations are naturally accommodated by the observational data.  In the figures, solid curves correspond to cubic spline interpolations of the constant maximum-probability values defined within each bin.

 Interestingly, the distributions are non gaussian. This is expected, as the mapping from the model-independent GP to the parameter space through the model is highly nonlinear.

Table \ref{tab_oml_4} shows the set of values of $\omega_{0\Lambda}$ at every emergent bin corresponding to an average redshift denoted as $z_{eff}$. The details of the emergent binning are presented in appendix \ref{app:f}.

 \begin{table}
 \centering
\begin{tabular}{|c|c|c|c|}
\hline
    $z_{eff}$ &  Most likely $\omega_{0\Lambda}$ & $z_{eff}$ &  Most likely $\omega_{0\Lambda}$ \\
    \hline
   0.24  & 0.223 & 1.91 & 0.214\\
   0.75 & 0.303 & 2.0 & 0.179\\
   1.05 & 0.321 & 2.06 & 0.071\\
   1.29 & 0.353 & 2.15 & 0.036\\
   1.54 & 0.393 & 2.23 & 0.357\\
   1.67 & 0.357 & 2.29 & 0.5\\
   1.79 & 0.321 & 2.35 & 0.464\\
   1.85 & 0.286 & & \\
   \hline
\end{tabular}
\caption{Emergent binning from the reconstruction of $\omega_{0\Lambda}$ using the $h(z)$ GP withing the four-free parameters subclass. The first column corresponds to effective redshifts while the second to the corresponding maximum posterior.}
\label{tab_oml_4}
\end{table}

Initially, we found that the Cubic Galileon prediction reconstruction is sensitive to the matter density of the Universe. Since this parameter is tightly constrained by a variety of cosmological observations, such as the primordial abundances of light elements predicted by Big Bang Nucleosynthesis (BBN), we assessed the viability of this subclass under these constraints. Accordingly, we allowed this parameter to vary rather than fixing it to the value inferred within the $\Lambda$CDM model. Fortunately, the reconstructed values of $\omega_{0m}$ remain consistent with the Planck-$\Lambda$CDM best-fit across the entire redshift range. 

Given that $\omega_{0m}$ is consistent with the Planck-$\Lambda$CDM constraints, we fix its value to the best-fit estimate in order to reduce the number of free parameters to three: $\alpha$, $\omega$ and $\omega_{0\Lambda}$. The main reason for this choice was that these parameters are the most relevant for studying the expansion of the universe as a gravitational outcome, which is in principle the main subject of our analysis. Namely, we expect that modifications of gravity in a considerable extent account for the accelerated expansion, so that the dynamics and kinematics of the scalar field controlled by $\alpha$ and $\omega$ would alter the required amount  of $\omega_{0\Lambda}$ at different redshifts in comparison with $\Lambda$CDM-like models. 

\begin{figure}
\centering
\includegraphics[scale = 0.6]{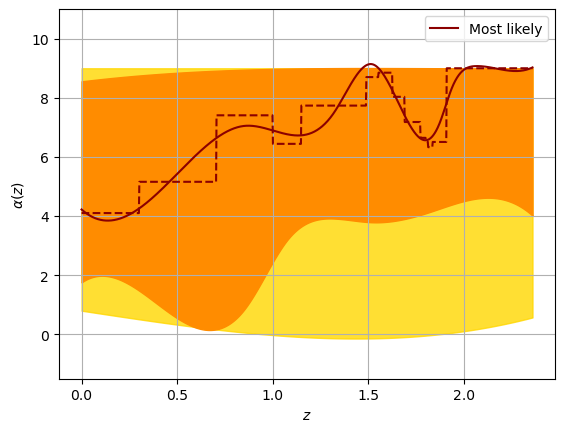}
\caption{\label{fig:alpha_h_3} Reconstruction of the parameter $\alpha$ using the  model-independent $h(z)$ GP in a three-dimensional parameter space. The $y$-axis has units of $M_{Pl}^2$. The most likely value is increasing with $z$ with an oscillatory behavior.}
\end{figure}

\begin{figure}
\centering
\includegraphics[scale = 0.6]{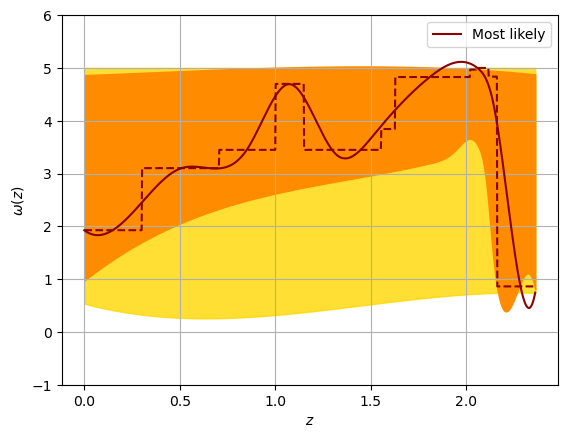}
\caption{\label{fig:omega_h_3} Reconstruction of the dimensionless parameter $\omega$ using the model-independent $h(z)$ GP in a three-dimensional parameter space. The $y$-axis is in logarithmic scale. The most likely value shows that the major modifications by this term happen in $z>2$.}
\end{figure}

\begin{figure}
\centering
\includegraphics[scale = 0.6]{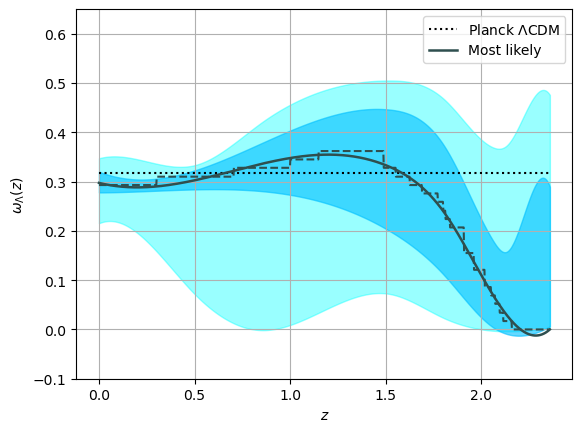}
\caption{\label{fig:omegal_h_3} Reconstruction of the physical density parameter $\omega_{\Lambda}$ using the model-independent $h(z)$ GP in a three-dimensional parameter space. The most likely value shows that $\omega_{0\Lambda}$ is not constant but decreasing with $z$ and reaching zero in $z>2$.}
\end{figure}

The resulting reconstructions are shown in figures \ref{fig:alpha_h_3}, \ref{fig:omega_h_3} and \ref{fig:omegal_h_3}, where the most likely value for each parameter as function of $z$ (dashed lines), once again these values give rise their binning from the posterior distribution of $\omega_{0\Lambda}$,  $z_{eff}$ and the corresponding most likely $\omega_{0\Lambda}$ are presented in table \ref{tab_oml_3}. 

\begin{table}
\centering
\begin{tabular}{|c|c|c|c|}
\hline
    $z_{eff}$ &  Most likely $\omega_{0\Lambda}$ & $z_{eff}$ &  Most likely $\omega_{0\Lambda}$ \\
    \hline
   0.15  & 0.293 & 1.82 & 0.224\\
   0.50 & 0.31 & 1.87 & 0.207\\
   0.85 & 0.328 & 1.93 & 0.155\\
   1.07 & 0.345 & 1.99 & 0.121\\
   1.31 & 0.362 & 2.03 & 0.086\\
   1.52 & 0.328 & 2.06 & 0.069\\
   1.58 & 0.31 & 2.08 & 0.052\\
   1.65 & 0.293 & 2.10 & 0.034\\
   1.73 & 0.276 & 2.13 & 0.017\\
   1.78 & 0.259 & 2.26 & 0.0\\
   1.80 & 0.241 & & \\
   \hline
\end{tabular}
\caption{Emergent binning from the reconstruction of $\omega_{0\Lambda}$ using the $h(z)$ GP within the three-free parameters subclass ($\omega_{0m}$ fixed). The first column corresponds to effective redshifts while the second to the corresponding most likely $\omega_{0\Lambda}$.}
\label{tab_oml_3}
\end{table}

Let us describe other reconstructions for this subclass corresponding to other dataset before presenting the analysis and discussion of these results.

\subsubsection{Reconstruction using the $h(z)$-BAO Gaussian Process}

In this section we present reconstructions for the Cubic Galileon subclass corresponding to the combined $h(z)$-BAO Gaussian Process reconstruction shown in figure \ref{fig:postDM}. 

For this analysis, the observable corresponds to the angular diameter distance $D_M$. Let us remind, that this quantity is measured for BAOs by using the sound horizon at the dragging epoch $r_d$ as standard ruler, therefore, the actual data corresponds to $D_M/r_d$. On one hand, we computed $D_M$ assuming a spatially-flat FLRW spacetime within the Cubic Galileon subclass. On the other hand, for computing the sound horizon, we used the model agnostic calibration for the sound horizon proposed at \cite{Liu_2024} which depends on $h_0$ derived from the model in question, $r_d^h = 100.83^{+0.99}_{-0.95}$Mpc$h_0^{-1}$. This calibration was made by using Gaussian Processes model-independent reconstructions on BAO-DESI data. The last approach is reasonable given that the Horndeski subclass described by action (\ref{galaction}) may not be valid at the early-universe because physics at those epochs may not be compatible with late time GW constraints.

By following the minimal approach used in the previous subsection, we reconstruct the posterior within the small parameter space given as  $(\alpha, \omega, \omega_{0\Lambda})$ which are relevant to analyze the gravitational sector of the Cubic Galileon subclass according to the model-independent $h(z)$-BAO Gaussian Process. Besides, $\omega_{0m}$ is fixed to Planck-$\Lambda$CDM best-fit, and we fixed initial conditions for $\phi_0$ and $\phi_0^\prime$ in the same way.

\begin{figure}
\centering
\includegraphics[scale = 0.6 ]{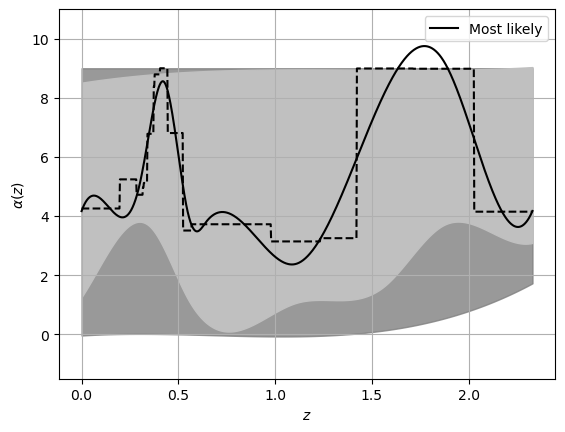}
\caption{\label{fig:alpha_BAO_3} Reconstruction of the Cubic Galileon self-coupling $\alpha$ according to the $h(z)$-BAO GP. Largest values of $\alpha$ arise at $z\sim 1.5-2.0$}
\end{figure}

\begin{figure}
\centering
\includegraphics[scale = 0.6]{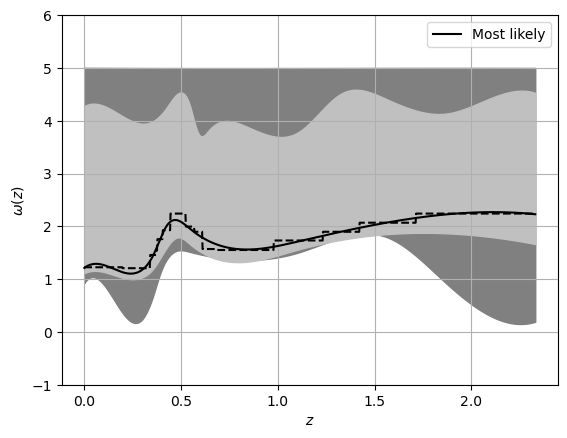}
\caption{\label{fig:omega_BAO_3} Reconstruction of the dimensionless parameter $\omega$ using the $h(z)$-BAO GP. The $y$-axis is in logarithmic scale. All along the redshift domain modifications of gravity by this term are most likely to be, being the major ones at low redshifts.}
\end{figure}

\begin{figure}
\centering
\includegraphics[scale = 0.6]{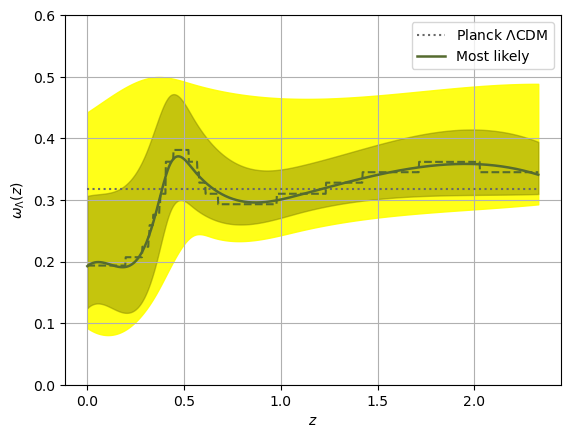}
\caption{\label{fig:omegal_BAO_3} Reconstruction of the physical density parameter $\omega_{\Lambda}$ using the $h(z)$-BAO GP. The dark energy density parameter varies with $z$ with the least value happening today at $z=0$, a bump happens at $z\sim0.5$, for greater values of redshift it decreases then increases again.}
\end{figure}

As it is customary, for a given $z$, we assign weights to each model by evaluating the (free-likelihood) posterior at each point in parameter space $\vec{\theta}$: 
\begin{equation}
P(\Vec{\theta},z|D_{\mathrm{comb}})\propto\exp{\left\lbrace-\frac{((D_{M}/r_d)_{\mathrm{pred}}(\Vec{\theta},z)-\overline{(D_M/r_d)} (z))^2}{2(\sigma(z))^2}\right\rbrace},
\end{equation}
where $(D_{M}/r_d)_\mathrm{pred}(\Vec{\theta},z)$ is the prediction for every model at each $z$, $\overline{(D_M/r_d)} (z)$ and $\sigma(z)$ correspond to the mean and standard deviation respectively of the model-independent Gaussian Process at a given $z$ corresponding to the combined dataset $D_{\mathrm{comb}}$ (BAO-$h(z)$). Let us recall that the resulting reconstructions for each parameter corresponds to the marginalized posterior which are shown in figures \ref{fig:alpha_BAO_3}, \ref{fig:omega_BAO_3} and \ref{fig:omegal_BAO_3}. As before,  the most likely values (which naturally arrange into bins) are plotted with dashed lines and solid lines correspond to a cubic spline of them. Also, $68\%$ and $95\%$ confidence regions are plotted as color bands. 

Intriguingly, all the reconstructions using either $h(z)$ or BAO-$h(z)$ Gaussian Process, show interesting differences. While, the tendency of the only-$h(z)$ phenomenological model clearly suggest that modifications to GR play a dominant role at large redshifts, in contrast, from the combined $h(z)$-BAO Gaussian Process the amount of dark energy needed for predicting $D_M$ exceeds the $\Lambda$CDM estimate at that redshifts. Although, large self-coupling effects seem to appear they have not relevant effect due to the increment of $\omega$ which promotes small amplitudes in variations of the galileon.    

For the purpose of determining the extent in which modifications to GR can replace dark energy, the binning for $\omega_{0\Lambda}$ arisen from its posterior is shown in table \ref{tab_oml_bao}.

\begin{table}
\centering
\begin{tabular}{|c|c|c|c|}
\hline
    $z_{eff}$ &  Most likely $\omega_{0\Lambda}$ & $z_{eff}$ &  Most likely $\omega_{0\Lambda}$ \\
    \hline
   0.09  & 0.193 & 0.54 & 0.362\\
   0.24 & 0.207 & 0.57 & 0.345\\
   0.29 & 0.224 & 0.59 & 0.328\\
   0.31 & 0.241 & 0.64 & 0.31\\
   0.33 & 0.259 & 0.82 & 0.293\\
   0.35 & 0.276 & 1.10 & 0.31\\
   0.37 & 0.293 & 1.32 & 0.328\\
   0.39 & 0.31 & 1.56 & 0.345\\
   0.42 & 0.362 & 1.87 & 0.361\\
   0.48 & 0.381 &  2.17 & 0.345\\
   \hline
\end{tabular}
\label{tab_oml_bao}
\caption{Emergent binning from the posterior distribution of $\omega_{0\Lambda}$ in the reconstructions using the $h(z)$-BAO GP. The data are the effective redshifts and the most likely value of $\omega_{0\Lambda}$ for each bin.}
\end{table}

We note that once again the posterior marginalized distributions are not gaussian, and they are in general different from the reconstructions obtained using the Gaussian Process reconstruction of $h(z)$ in the last section. The difference is no surprise because in both cases different physical processes affect the observable, recalling that in the case of the BAO measurements the sound horizon is playing a role invoking early-universe physics or model-agnostic assumptions like the one used in this work, whilst the Hubble parameter measurements depend only on the late-time cosmology.

\section{\label{sec:discussion}Discussion of the results}

\subsection{Model-independent reconstructions}

The model-independent Gaussian Process reconstruction of $h(z)$ is inconsistent with the $\Lambda$CDM best-fit prediction at the $68\%$ confidence level around redshifts $z\sim2$. This discrepancy is even more pronounced for the predictions of constant-$\omega$ Brans--Dicke models, suggesting that additional modifications to gravity may be required for these theories to accurately describe the Hubble parameter in this redshift regime. In contrast, at intermediate redshifts, several constant-$\omega$ Brans--Dicke models remain consistent with the Gaussian Process reconstruction, whereas the $\Lambda$CDM best-fit prediction lies outside the $1\sigma$ confidence region.

The phenomenological Gaussian Process reconstruction of $D_M(z)$ based on the $h(z)$--BAO dataset is nearly indistinguishable from the $\Lambda$CDM best-fit prediction for redshifts $z\gtrsim1.5$. At lower redshifts, however, the two reconstructions begin to differ, with the largest discrepancies appearing for $z\lesssim0.5$. This behavior is a consequence of the lack of observational constraints on $D_M(z)$ in this redshift range. As a result, the Gaussian Process extrapolates the reconstruction using the covariance between the observed data and the prediction points.

To incorporate observational information across the entire redshift range, we considered the combined dataset. The resulting Gaussian Process reconstruction yields a phenomenological model that is in better agreement with the observations than the $\Lambda$CDM best-fit, as indicated by the $\chi^2$ analysis resumed in Table \ref{tab:chisq_bao}. 

\subsection{Brans--Dicke subclass} 
Predictions for the Hubble parameter within this subclass are even more overestimated than those of the $\Lambda$CDM model, with the discrepancy becoming increasingly pronounced at higher redshifts. According to the reconstruction of $\omega$ at the $68\%$ confidence level, the most favored cosmological model within the redshift interval $z \in (0.4,1.7)$ is from the Brans--Dicke subclass rather than $\Lambda$CDM. Moreover, the corresponding preferred values of $\omega$ are consistent with the constraints previously obtained from Planck and other cosmological datasets. It is worth emphasizing that the model-independent Gaussian Process reconstruction statistically favors the prediction of the Brans--Dicke subclass over that of $\Lambda$CDM. In this sense, BD models provide a better description of the Hubble expansion rate than $\Lambda$CDM within the previously mentioned redshift interval.

\subsection{Cubic galileon subclass}

We begin by analyzing the reconstructions of the Cubic Galileon's free functions obtained from deterministic ensembles of models with either four or three free parameters, constructed from the model-independent Gaussian Process reconstruction of $h(z)$. As expected, the multidimensionality of the parameter space of these ensembles leads to broad confidence regions for the reconstructed free functions, even when the uncertainty of the model-independent reconstruction is small.

Regarding the reconstructions of $\alpha(z)$, in both cases, the most likely values oscillate inside the range $\alpha\in(2,9)$ across the redshift range, which are consistent with estimates associated with a cosmological-scale Vainshtein radius. It is important to note that at $z\gtrsim 1$, $\alpha=0$ lies within the $95\%$ confidence region, indicating that the data are consistent with the absence of modifications to gravity at these epochs. A notable feature of the $\alpha$ reconstruction within the 3-parameter instance, is that the most-likely values shift above(below) the average value for high (low) redshift, compensating for the lack of freedom in $\omega_{m0}$. In contrast, $\alpha$ most-likely values oscillate around the average within the 4-parameter instance.

The overall behavior of reconstructions of $\omega(z)$  in both cases, indicate that the most significant deviations from GR, driven by a Brans--Dicke-like enhancement through variations of the gravitational coupling, are more likely at high redshifts. In contrast, at low redshift, larger values of $\omega$ are preferred, which are associated with the GR limit of constant-$\omega$ Brans--Dicke models according to cosmological constraints. More specifically, in the four-parameter ensemble, relatively small values of $\omega$ laying in the range $(0,10)$ occur at $z>1.5$. Such small values imply substantial variations in the gravitational coupling, corresponding to a strong Brans--Dicke enhancement. In the three-parameter ensemble, this Brans--Dicke enhancement is shifted toward higher redshifts.
In the former ensemble, the data favor the GR limit of this Brans--Dicke subclass within the interval  $1.0<z<1.5$. In the latter ensemble, however, this behavior extends over a broader redshift range. 
Interestingly, in the three-parameter case, this behavior appears to be correlated with the dynamical modifications encoded in the free function $\alpha$, whose magnitude increases over the same redshift range.

Let us recall that, within the four-parameter ensamble, the reconstruction of $\omega_{0m}$ is consistent with the Planck-$\Lambda$CDM constraint reported by the Planck Collaboration. Notably, this free function is the most tightly constrained among those reconstructed in our analysis. The decision to fix this parameter within the three-parameter ensamble is motivated not only by its agreement with the Planck constraint, but also by theoretical considerations. In the Cubic Galileon subclass, the scalar field is not expected to couple directly to matter due to the choice of the Jordan conformal frame. Consequently, the scalar-field dynamics, governed by the free functions $\alpha$ and $\omega$, are not expected to significantly affect the matter density parameter.

Finally, the reconstruction of $\omega_{0\Lambda}$ in both ensembles, indicates a well-defined redshift-dependent evolution of their most probable value.  These most likely values remain below the Planck-$\Lambda$CDM estimate at low redshifts, increasing monotonically until reaching this reference value. Afterwards, at intermediate redshift they remain above it and reach their maximum around $z\sim1.3-1.5$. At higher redshift ($z\gtrsim1.5$), most-likely $\omega_{0\Lambda}$ decrease approaching zero around $z\sim2$. 
As expected, the non-linear dynamics of the scalar field contribute to the accelerated expansion of the Universe, thereby changing the dark energy density required compared with the standard $\Lambda$CDM model. The dominant contribution of these modifications to GR becomes increasingly significant toward higher redshifts.
The fact that the most probable value of $\omega_{0\Lambda}$ approaches zero for  $z>2$ suggests that the observations favor significant modifications of GR at these epochs. This trend can be also observed in the $\alpha$ and $\omega$ reconstructions and appears to be closely connected with the deviation of the model-independent Gaussian Process reconstruction of $h(z)$ from the $\Lambda$CDM prediction at these redshifts.
Remarkably, for these reconstructions of $\omega_{0\Lambda}$, the uncertainty is considerably reduced compared with that of $\alpha(z)$ and $\omega(z)$, especially within the three-parameter ensemble.

\paragraph{Reconstructions from $h(z)$-BAO GP} 
 
In contrast with reconstructions arisen  from $h(z)$ data, from the combined $h(z)$-BAO Gaussian Process the reconstruction of $\alpha$ has the most-likely values different from zero across the redshift domain. Two ranges of redshift can be recognized, where the largest modifications to GR appear: $z\sim 0.4$ and $z\in(1.4, 2.0)$. For intermediate redshift values, the most likely $\alpha$ take smaller values indicating less important modifications to gravity associated to the galileon dynamics.

Regarding the reconstruction of $\omega$, the most probable values of this free function lie in the range $\omega \sim 10\text{--}10^2$. Cosmological constraints indicate that, for $\omega \gtrsim 890$, the predictions of Brans--Dicke theory become observationally indistinguishable from those of $\Lambda$CDM. Therefore, the reconstructed values correspond to deviations from General Relativity that remain relevant throughout the entire redshift range considered. The most prominent modifications of this kind occur at the minimum located at ($z=0$). The highest values of the most probable $\omega$, occurring at $z \lesssim 0.5$ and $z \sim 1.5\text{--}2$, appear to be associated with strong Galileon nonlinearities at these redshifts, as reflected in the reconstruction of $\alpha(z)$. Interestingly, these nonlinearities arise in regions where the Brans--Dicke enhancement is suppressed and the most probable value of $\omega_{\Lambda0}$ increases above the Planck-$\Lambda$CDM estimate, possibly compensating for the combined effects of the Galileon nonlinearities and the Brans--Dicke enhancement.  A more detailed investigation of the interplay between these distinct modifications is left for future work. Nevertheless, the reconstruction provides evidence that these features of the model may be required to simultaneously reproduce the BAO and $h(z)$ observations.

Notice that $\omega_{\Lambda0}$ is the best-constrained parameter in these reconstructions, showing significantly smaller uncertainties than the remaining free functions. Additionally, around $z=0$, the most probable value of $\omega_{\Lambda0}$ reaches its lowest value, corresponding to the largest Brans--Dicke-enhancement-like deviations from GR, with $\omega \sim 10$. At the same time, the reconstruction favors a nonzero value of $\alpha$ near $z=0$, suggesting that the data favor dynamical modifications of gravity rather than a larger dark energy contribution.

The evident differences between the reconstructions from $h(z)$ and BAO-$h(z)$ arise because cosmic expansion and the growth of large-scale structures, as traced by BAO, are governed by different physical processes. Taken separately or together, these reconstructions provide a powerful, non-parametric approach to unveiling the structure of the free functions of Horndeski cosmological models directly from observations, without imposing theoretically motivated functional forms.

\section{\label{sec:conclusions}Conclusions}
In this work, the free functions of the Horndeski subclass compatible with the GW170817 constraints, as described by the action (\ref{galaction}), are reconstructed using the Gaussian Process technique according to observations characterizing the expansion of the universe. Specifically, Hubble parameter measurements from CC and BAO-BOSS-SDSS, and angular diameter distances from BAO-DESI-DR2. Firstly we obtained Gaussian Process reconstructions of $h(z)$ and $D_M(z)$ across a redshift range given by $z\in(0,2.3)$. The most-likely values of these model-independent reconstructions define phenomenological curves that describe each observable. Remarkably, these curves achieve a better goodness-of-fit than the $\Lambda$CDM model.

We carried out reconstructions of free functions of the Horndeski subclass considering two cases of the action: the Brans--Dicke and the Cubic Galileon subclasses. We first reconstructed the single-parameter Brans--Dicke subclass for pedagogical purposes, which describes a free, non-canonical scalar field non-minimally coupled to gravity which acts as a dynamical gravitational coupling. In this case, a cosmological constant is included to account for the accelerated expansion of the Universe since the model corresponds to a simple extension of GR. We then reconstructed the free functions of the Cubic Galileon subclass, which extends the former model by incorporating dynamical Cubic Galileon like self-interactions of the scalar field that give rise to self-accelerating solutions. 

 Let us summarize what is meant by the reconstruction of a free parameter within the Horndeski subclass over a specified redshift range. Given an observational dataset, a model-independent Gaussian Process is constructed, which may be interpreted as a redshift-parametrized Gaussian distribution consistent with the observational data. At each redshift $z$, we evaluate the posterior distribution in the parameter space by weighting the model predictions according to this model-independent reconstruction. The reconstruction of each free parameter is then obtained by marginalizing the resulting posterior distribution. It is important to stress that the reconstructed Horndeski subclasses are fundamentally different from deterministic models with constant parameters specified {\it a priori}. Each reconstructed subclass corresponds to a family of theories characterized by redshift-dependent free functions rather than constant parameters. In contrast of a parametric and deterministic approach, within this framework such free functions are statistically described by the reconstruction, and their resulting probability distributions are not necessarily Gaussian. It is remarkable that, in any case, the most likely values in the reconstruction remain constant across the redshift bins which are naturally accommodated by the observational data.

In the case of the Brans--Dicke subclass, we used only the $h(z)$ reconstruction to illustrate the methodology. 

To reconstruct the free functions of the Cubic Galileon subclass, we employed the Gaussian Process reconstructions of $h(z)$ and of the combined $h(z)$--BAO. Since the predictions for $h(z)$ within the Cubic Galileon model are sensitive to the value of $\omega_{0m}$, we first performed the reconstruction allowing this parameter to vary together with $\alpha$, $\omega$, and $\omega_{0\Lambda}$. The reconstructed value of $\omega_{0m}$ was found to be consistent with that inferred from the Planck $\Lambda$CDM analysis. We therefore carried out a second reconstruction using a reduced parameter space, with $\omega_{0m}$ fixed to its $\Lambda$CDM value. Using the $h(z)$ reconstruction, the most likely values of the dark energy density parameter $\omega_{0\Lambda}$ vary smoothly with redshift. Remarkably, values close to zero are preferred at high redshifts, where the reconstructed functions $\alpha$ and $\omega$ suggest significant modifications to gravity. In contrast, for the $h(z)$--BAO Gaussian Process reconstruction, the free functions differ from those obtained previously and lead to a less conclusive physical interpretation. This is likely due to two main factors: (1) the absence of a well-established description of early-Universe physics for the Cubic Galileon model, which was circumvented by adopting a model-independent calibration of the sound horizon; and (2) the computational difficulty of accurately evaluating the quantity $D_M(z)$ within the Cubic Galileon model, owing to the integration of its highly nonlinear cosmological field equations.

In conclusion, the Gaussian Process reconstructions presented in this work provide a distribution of the free functions $\alpha(z)$ and $\omega(z)$ describing the Horndeski subclasses, that is consistent with Hubble parameter measurements and the angular diameter distance inferred from BAO observations. Furthermore, these reconstructions consistently favor a redshift-dependent dark energy density parameter. 

Although the Brans--Dicke subclass is unable to reproduce the observed cosmic expansion history without invoking dark energy at all redshifts, it's interesting that the reconstruction of $\omega$ suggest an enhancement of the variation of the gravitational coupling to matter 
at intermediate redshifts. In contrast, the Cubic Galileon subclass is capable of reproducing the Hubble expansion rate without requiring dark energy at high redshifts $z>2$, where the largest modifications to gravity associated with $\alpha$ and $\omega$ appear. However, simultaneously fitting the BAO and $h(z)$ data requires a significant amount of dark energy. The reconstructions obtained from this combined analysis reveal an intriguing interplay between the nonlinear Galileon dynamics, the Brans--Dicke enhancement, and a redshift-dependent dark energy component.

\begin{acknowledgments}
MCMM acknowledges SECIHTI-M\'exico for the PhD studies scholarship. AAAL work has been partially supported by S.N.I.I. (SECIHTI-M\'exico). MCMM and AAAL acknowledge VIEP-BUAP for supporting this work by means of the "Proyectos VIEP 2026" program with I.D. 00844-PV/2026.
\end{acknowledgments}

\appendix

\section{\label{app:a}Field equations in Horndeski theory}
We derived the field equations for the action (\ref{galaction}) by means of the variational principle. Taking the variation with respect to $g^{\mu\nu}$ we obtained the tensorial equation
\begin{eqnarray}
    \label{tensoreqGal}
        8\pi GT_{\mu\nu}&=&G_4(\phi)(R_{\mu\nu}-\frac{1}{2}Rg_{\mu\nu})+G_4^{\prime}(\phi)(\Box\phi g_{\mu\nu}-\nabla_{\mu}\nabla_{\nu}\phi)-G_4^{\prime\prime}(\phi)(2Xg_{\mu\nu}+\nabla_{\mu}\nabla_{\nu}\phi)\nonumber\\
        &&+G_{3,X}\left(\frac{1}{2}\Box\phi\nabla_{\mu}\phi\nabla_{\nu}\phi+\nabla_{\mu}X\nabla_{\nu}\phi-\frac{1}{2}\nabla_{\sigma}\phi\nabla^{\sigma}Xg_{\mu\nu}\right)+G_{3,\phi}(\nabla_{\mu}\phi\nabla_{\nu}\phi+Xg_{\mu\nu})\nonumber\\
        &&-\frac{1}{2}G_2 g_{\mu\nu}-\frac{1}{2}G_{2,X}\nabla_{\mu}\phi\nabla_{\nu}\phi.
\end{eqnarray}
And varying the action with respect to the scalar field, we obtained, after eliminating $R$ with the trace of (\ref{tensoreqGal}), the scalar equation
    \begin{eqnarray}
        \label{scalareqGal}
    \left(\frac{G_4^{\prime}}{G_4}\right)8\pi T&=&G_{2,\phi}-2XG_{2,X,\phi}-G_{2,X,X}\nabla_{\alpha}X\nabla^{\alpha}\phi+\left(G_{2,X}-G_{3,\phi}+2XG_{3,X,\phi}-G_{3,X,X}\nabla_{\alpha}X\nabla^{\alpha}\phi\right)\Box\phi\nonumber\\
    &&-G_{3,X}(\Box\phi)^2-G_{3,X}\nabla^{\alpha}\phi\nabla_{\alpha}\Box\phi-\Box G_3+3\frac{(G_4^{\prime})^2}{G_4}\Box\phi-6X\frac{G_4^{\prime}G_4^{\prime\prime}}{G_4}\nonumber\\
    &&-\frac{G_{3,X}G_4^{\prime}}{G_4}(\Box\phi X+\nabla^{\alpha}X\nabla_{\alpha}\phi)-2\frac{G_{3,\phi}G_4^\prime}{G_4}X-2\frac{G_2 G_4^\prime}{G_4}+\frac{G_{2,X}G_4^\prime}{G_4}X.  
    \end{eqnarray}

\section{\label{app:b}Gaussian processes}

The Gaussian Process reconstruction technique ~\cite{GP} is a Bayesian statistical method to perform regression or classification over a set of observations. In the case of regression, given a dataset $\lbrace(\textbf{x}_i, y_i)|_{i=1}^n\rbrace$ the Gaussian Process aims to find the predictive model $f(\textbf{x}_i)$ for all $i$-element in the data set, as well as a region of confidence built assuming that exists a covariance function for every pair of  $(f(\textbf{x}_i), f(\textbf{x}_j))$ such that
\begin{equation}
    \cov(f(\textbf{x}_i), f(\textbf{x}_j))=k(\textbf{x}_i, \textbf{x}_j)
\end{equation}
holds, i.e. the covariance between outputs is a function of the inputs. The covariance function or kernel $k$ can be chosen from a wide range of possibilities, the most widely used is the \textit{squared exponential kernel}  
\begin{equation}
\label{sqexp}
    k(\textbf{x}_i, \textbf{x}_j) = \sigma_f^2\exp{\left\lbrace-\frac{|\textbf{x}_i- \textbf{x}_j|^2}{2l^2}\right\rbrace}.
\end{equation}
 It depends on two hyperparameters $\sigma_f$ and $l$ that characterize the shape of the function $f(\textbf{x)}$. $\sigma_f$ controls the typical change in the direction $y$, and $l$ the distance in the direction $\textbf{x}$ needed to obtain a significant change in $f(\textbf{x)}$. This kernel is useful for reconstructing derivatives of a function, because of its infinite differentiability. Another commonly used kernel is the \textit{Matern kernel}

\begin{equation}
\small
    k(\textbf{x}_i, \textbf{x}_j) = \frac{\sigma_f^2}{\Gamma(\nu) 2^{\nu-1}}\left(\frac{\sqrt{2\nu}}{l}|\textbf{x}_i- \textbf{x}_j|\right)^\nu K_\nu \left(\frac{\sqrt{2\nu}}{l}|\textbf{x}_i- \textbf{x}_j|\right),
    \label{matern}
\end{equation}
\vspace{2mm}

\noindent which is a generalization of the squared exponential but with an additional hyperparameter $\nu$, which controls the smoothness of the function $f(\textbf{x})$, as $\nu\rightarrow\infty$ the squared exponential is recovered; $K_\nu$ is a modified Bessel function of the second kind and $\Gamma$ is the gamma function. These two kernels are used in this work.

The advantage of using Gaussian Processes over other Bayesian methods such as Markov Chain Monte Carlo (MCMC) ~\cite{speagle2020conceptualintroductionmarkovchain} or the nested sampling algorithm ~\cite{Buchner_2023} lies in its non-parametric approach, called the \textit{function space view}. Instead of fitting a parametric model, the Gaussian Process based on a chosen kernel generates a posterior distribution 
\begin{eqnarray}   \boldsymbol{f^*}|\boldsymbol{X^*},\boldsymbol{X},\boldsymbol{y}\sim\mathcal{GP}(\bar{\boldsymbol{f^*}},\cov(\boldsymbol{f^*})),
\end{eqnarray}
statistically consistent with the observational data. Here, $\boldsymbol{X}$ and $\boldsymbol{X^*}$ are matrices whose columns contain the input vectors $\mathbf{x}_i$ from the observations and the prediction points $\mathbf{x}_j^*$ defining the reconstruction domain, respectively, while $\boldsymbol{y}$ denotes the vector of observed outputs. The predictive distribution is fully characterized by its mean function and covariance matrix, given by
\begin{eqnarray}  \label{fstar}  \boldsymbol{\bar{f^*}}=\boldsymbol{\mu^*}+K(\boldsymbol{X^*},\boldsymbol{X})[K(\mathbf{X},\mathbf{X})+C]^{-1}(\mathbf{y}-\boldsymbol{\mu})
\end{eqnarray}
and
\begin{eqnarray}   \label{covfstar} \cov(\boldsymbol{f^*})=K(\boldsymbol{X^*},\boldsymbol{X^*})-K(\boldsymbol{X^*},\boldsymbol{X})[K(\boldsymbol{X},\boldsymbol{X})+C]^{-1}K(\boldsymbol{X},\boldsymbol{X^*}). \nonumber
\end{eqnarray}
 Here, $K$ denotes the covariance matrix, whose elements are given by Eqs.~(\ref{sqexp}) or (\ref{matern}); $C$ is the covariance matrix of the observations; $\boldsymbol{\mu}$ is the mean vector of the observations; and $\boldsymbol{\mu^*}$ is the prior mean of $\boldsymbol{f^*}$, which can be safely assumed to be constant. The predictive variance of $\boldsymbol{f^*}$ is given by the diagonal elements of $\mathrm{cov}(\boldsymbol{f^*})$.

To fully specify the Gaussian Process, the kernel hyperparameters must be determined. For the squared exponential kernel, Eq.~(\ref{sqexp}), these are the signal amplitude $\sigma_f$ and the characteristic length scale $l$, while the Matérn kernel, Eq.~(\ref{matern}), additionally depends on the smoothness parameter $\nu$. These hyperparameters can be inferred by maximizing the log marginal likelihood,
\begin{eqnarray}
\ln{\mathcal{L}}=-\frac{1}{2}(\mathbf{y}-\boldsymbol{\mu})^{T}[K(\mathbf{X},\mathbf{X})+C]^{-1}(\mathbf{y}-\boldsymbol{\mu})-\frac{1}{2}\ln{|K(\mathbf{X},\mathbf{X})+C|}-\frac{n}{2}\ln{2\pi},
\end{eqnarray}
or more generally, by marginalizing over the hyperparameter space using optimization methods such as gradient descent or MCMC.

Gaussian Processes have been widely used in cosmology to reconstruct a variety of quantities and observables, including the dark energy equation of state~\cite{Vel_zquez_2024}, cosmographic parameters~\cite{2023EPJC...83..374L}, and predictions of gravitational theories beyond General Relativity \cite{Gadbail_2024, Ren_2021, Ren_2022}. Several software packages have also been developed to implement Gaussian Process regression, including the \texttt{scikit-learn} library~\cite{scikit-learn} and the \texttt{GaPP} code \cite{GP}. Both are employed in the present work.

\section{\label{app:c}Backgound cosmology equations for the Brans--Dicke theory}
The system of ordinary first order equations corresponding to the constant-$\omega$ Brans--Dicke theory in a FLRW spacetime, which describes the dynamics of elements in the ensemble for the first Horndeski subclass reads
\begin{equation}
    \label{phiBD}
    \phi^\prime = \left\lbrace\pm\sqrt{\left(\frac{6}{\omega}+\frac{9}{\omega^2}\right)\phi^2-\frac{6}{\omega h^2}\left(\omega_{0m}(1+z)^3+\omega_{0\Lambda}\right)\phi}-\frac{3}{\omega}\phi\right\rbrace
    \times (1+z)^{-1},
\end{equation}

\begin{equation}
    \label{hBD}
    h^\prime = \frac{3}{(4\omega+6)\phi h}\left[\omega_{0m}(1+z)^2+(1-2\omega)\omega_{0\Lambda}(1+z)^{-1}\right]
    +\frac{\omega}{4}h(1+z)\left(\frac{\phi^\prime}{\phi}\right)^2+\frac{3h}{2(1+z)}.
\end{equation}

\section{\label{app:d}Backgound cosmology equations for the cubic galileon theory}
For the Cubic Galileon subclass reconstruction the complete system of equations (\ref{firstgaleq}, \ref{secondgaleq}, \ref{thirdgaleq}) must be solved. Similarly in the Brans--Dicke case, a change of variable  $t\rightarrow z$ is performed, the energy density and pressure are expressed in terms of the physical density parameters and the normalized Hubble parameter $h(z)\equiv H(z)/(100km/seg/Mpc)$ is used as variable. In this case the normalized Hubble constant $h_0$ will be a derived parameter rather than a free one. Within the Brans--Dicke case, the approach was different because the system of equations can be reduced to a first order system and $h_0$ enters as the initial condition for the $h(z)$ variable. In constrast, for the Cubic Galileon the second order system of equations can not be reduced and then it was solved using $\phi$ and its derivative with respect to $z$, $\phi^\prime$ as dynamical variables. After changing the variables in (\ref{firstgaleq}), (\ref{secondgaleq}) and (\ref{thirdgaleq}) the following equations are obtained

    \begin{eqnarray}
        \label{galzeq1}
        3\sum_{i}\omega_{0i}(1+z)^{3(1+w_i)}&=&3h^2\phi-3h^2(1+z)\phi^\prime-\frac{\omega}{2\phi}h^2(1+z)^2\phi^{\prime 2}\nonumber\\
        &&+\frac{3 \tilde{\alpha}}{8}h^4(1+z)^3\left(\frac{\phi^\prime}{\phi}\right)^3 -\frac{3\tilde{\alpha}}{16}h^4(1+z)^4\left(\frac{\phi^\prime}{\phi}\right)^4.
    \end{eqnarray}
This corresponds to the generalized Friedmann equation within this Cubic Galileon model. By evaluating it in $z=0$, the constraint equation is derived which corresponds to a fourth degree polynomial equation for $h_0$ given by
\vspace{2mm}
\begin{eqnarray}
    &&0= h_0^4\left[\frac{3\alpha}{8}\phi_0^{\prime3}\left(\phi_0-\frac{1}{2}\phi^\prime_0\right)\right] + h_0^2\left[3\phi_0^4(\phi_0-\phi^\prime)-\frac{\omega}{2}\phi_0^{\prime 2}\phi_0^3\right]
    -3\phi^4\sum_i \omega_{0i}(1+z)^{3(1+w_i)}. \nonumber\\
    &&
\end{eqnarray}

\noindent Therefore, $h_0$ depends on the combination of $\alpha$, $\omega$, the physical density parameters, and the values of the scalar and its derivative today.

The equation arising from the spatial component of the tensor equation (\ref{tensoreqGal}) in terms of $z$ reads:

    \begin{eqnarray}
        \label{galzeq2}  &&3\sum_iw_i\omega_{0i}(1+z)^{3(1+w_i)}=\phi[-3h^2+2(1+z)hh^\prime]+\phi^\prime[2h^2(1+z)-hh^\prime(1+z)^2]-h^2(1+z)^2\phi^{\prime\prime}\nonumber\\
        &&-\frac{\omega}{2\phi}h^2\phi^{\prime2}(1+z)^2
        +\frac{\tilde{\alpha}}{8\phi^3}h^3\phi^{\prime2}(1+z)^3\left[\phi^\prime(h^\prime(1+z)+h)-\frac{3}{2\phi}h(1+z)\phi^{\prime2}+h(1+z)\phi^{\prime\prime}\right]. 
    \end{eqnarray}
  
 Finally, the scalar equation (\ref{scalareqGal}) in terms of $z$ is given by

    \begin{eqnarray}
        \label{galzeq3}
        &&3\sum_i\omega_{0i}(3w_i-1)(1+z)^{3(1+w_i)}=(2\omega+3)\left[h(1+z)(2h-h^\prime(1+z))\phi^\prime-h^2(1+z)^2\phi^{\prime\prime}\right] \nonumber\\
       && -\frac{3\tilde{\alpha}}{4\phi^2}\left[-3h^3h^\prime(1+z)^3\phi^{\prime2}+h^4(1+z)^2\phi^{\prime2}+\frac{2}{\phi}h^3(1+z)^3\left(h+\frac{3}{4}h^\prime(1+z)\right)\phi^{\prime3}\right. \nonumber\\
        &&\left.-\frac{5}{2\phi^2}h^4(1+z)^4\phi^{\prime4}-2h^4(1+z)^3\left(1-\frac{3}{4\phi}(1+z)\phi^\prime\right)\phi^\prime\phi^{\prime\prime}\right].
    \end{eqnarray}
The prime indicates derivative with respect to $z$ and $\tilde{\alpha}\equiv\alpha(100km/sec/Mpc)^2$, corresponds to the kinetic-braiding coupling parameter in units of $M_{Pl}^2$. Let us recall that the original $\alpha$ was dimensionless.

\section{\label{app:e}Priors and initial conditions for the cubic galileon}
 In Galileon models, the most robust constraints are obtained when all, or at least most, of the model parameters are let to vary freely, thereby enabling a thorough exploration of the possible parameter degeneracies~\cite{Barreira_2014, Renk_2017}. In this analysis, four parameters are allowed to vary freely, namely $\alpha$, $\omega$, the physical density of matter $\omega_{0m}$ and  cosmological constant $\omega_{0\Lambda}$. The latter is included to quantify the extent to which the Cubic Galileon model can alleviate the need for a cosmological constant.
The initial condition for the scalar field is fixed to $\phi_0 = 0.989$, in agreement with cosmological constraints on the gravitational constant~\cite{Ballardini_2022}, since the present-day value of the scalar field determines the effective gravitational coupling through $\phi_0 = G_{\rm eff}/G_N$. The initial condition for its derivative is set to $\phi_0^\prime = 1\times10^{-4}$, corresponding to the fiducial Brans--Dicke model analyzed previously. This choice is motivated by the fact that Brans--Dicke theory can be regarded as an approximation to Horndeski theory on cosmological scales~\cite{Avilez:2013}.

Uniform priors are adopted for all parameters in order to broadly explore the parameter space defined by the Cartesian product of $\alpha\in[0,10]$, $\omega\in\log([0,4])$, $\omega_{0m}\in[0,0.5]$, and $\omega_{0\Lambda}\in[0,0.5]$. The prior on $\alpha$ is estimated using Eq.~(\ref{alpha_funlambda}). The prior on $\omega$ is chosen similarly to that adopted in the Brans--Dicke analysis, while the priors on the physical density parameters are selected to satisfy $0\leq\omega_{0i}<h_0^{-2}$.

\section{\label{app:f}Binning selection}
In all the reconstructions presented here, the binning scheme was selected according to the one obtained from the $\omega_{0\Lambda}$ reconstruction. This choice is motivated by the goal of analyzing the accelerated expansion of the Universe within the Horndeski framework. While in the standard $\Lambda$CDM, it is assumed to be constant, within the Cubic Galileon subclass, $\omega_{0\Lambda}$ takes different values in each redshift bin. This behavior indicates a possible redshift dependence of the dark energy density parameter.

We emphasize that this binning is not arbitrary; rather, it is determined by the most-likely values of $\omega_{0\Lambda}$, which vary with each effective redshift,
\begin{equation}
z_{\rm eff}=\frac{z_i+z_{i+1}}{2},
\end{equation}
resulting in a distinct value of $\omega_{0\Lambda}$ for each redshift interval $(z_i,z_{i+1})$.

\bibliographystyle{JHEP}
\bibliography{ref}

\end{document}